\documentclass{article}     
\usepackage{graphicx}
\usepackage{epstopdf}
\title{A systematic study of longitudinal and transverse helicity amplidutes in the hypercentral
Constituent Quark Model}  
\author{E. Santopinto\\
I.N.F.N., Sezione di Genova\\
\\
M.M. Giannini\\
Dipartimento di Fisica dell'Universit\`a di Genova\\
and \\
I.N.F.N., Sezione di Genova\\}

\date{}
\begin{document}             

\maketitle         

\begin{abstract}
We report in a systematic way the predictions of the non-relativistic hypercentral Constituent Quark model
for the electromagnetic excitations of baryon resonances. The longitudinal and transverse helicity amplitudes are calculated 
with no free parameters for fourteen resonances, both for proton and neutron. The calculations lead to an overall fair description of data, 
specially in the medium $Q^2$ range, where quark degrees of freedom are expected to dominate.
\end{abstract}
%\pacs{12.39.Jh Non relativistic quark model-14.20.Gk Baryon resonances (S=C=B=0)-13.40.Gp Electromagnetic form
%factors}

ÅÅ

\section{Introduction}

Various Constituent Quark Models (CQM) \cite{morp65} have been proposed in the recent past
 after the pioneering work of Isgur and Karl (IK) \cite{ik}. Among
them we quote the relativized Capstick-Isgur (CI) model
\cite{ci}, the algebraic approach (BIL) \cite{bil}, the hypercentral CQM
(hCQM) \cite{pl,nc,es}, the chiral Goldstone Boson Exchange model (GBE)
\cite{olof,ple} and the Bonn instanton model  (BN) \cite{bn}.

The ingredients of the models are quite different, but they have a simple
general structure, since they can be split into a spin-flavour independent
part $V_{inv}$, which is $SU(6)$-invariant and contains the confinement
interaction, and a $SU(6)$-dependent part $V_{sf}$, which contains spin
and eventually flavour dependent interactions, in agreement with the
prescription provided by the early Lattice QCD calculations \cite{wil}. For the latter,  the hyperfine 
interaction is often used \cite{deru}. 
They are all able
to reproduce the baryon spectrum, which is the first quantity to be described but a real test 
of the models is provided by their systematic and consistent application
to the description of other physical quantities of the nucleon.
In this respect it is interesting to see to which extent and how systematically the various CQM have been used; one should not forget that in many cases the calculations referred to as a CQM one are actually performed using a simple h.o. wave function for the internal quark motion either in a non relativistic (HO) or relativistic framework (rHO).

The photocouplings for the excitation of the baryon resonances have been  calculated in various models, among others we quote HO \cite{cko}, IK \cite{ki}, CI  \cite{cap}, BIL  \cite{bil}, hCQM  \cite{aie} (for a comparison among these and other
approaches see e.g. \cite{aie,cr2}). The calculations are in general able to reproduce the overall trend, but the strength is systematically lower than the data; such similarity of results coming from quite different models can be ascribed to the common $SU(6)$ structure quoted above.

In many cases the models hav been applied to the  description of the elastic nucleon form factors. The algebraic method by BIL \cite{bil,bil2} has been used, assuming a definite charge distribution along the string connecting quarks, while the CI model is the basis of a light front calculation by the Rome group \cite{card_ff}, obtaining a good description of data, provided that quark form factors are introduced.
As for the hCQM, it has been firstly applied in the non relativistic version with Lorentz boosts \cite{mds,rap}, showing that the recently observed \cite{rap_exp} behavior of the ratio between the charge and magnetic form factors of the proton may be ascribed to relativistic effects; the hCQM has been reformulated relativistically in a point form approach, the resulting elastic nucleon form factors are quite good and further improved by the introduction of intrinsic quark form factors \cite{ff_07,ff_10}. A quite good description of the elastic form factors is achieved also using the GBE \cite{wagen,boffi} and the BN \cite{mert} models, both being fully relativistic.

The calculation of the elastic form factors allows to understand to which extent the ground state is under control, whereas the study of the $Q^2$ behavior of the excitation to the baryon resonances provides a sensible test of both the energy and the short range properties of the quark structure. This fact motivates the attention which has been devoted to the electromagnetic transition form factors (helicity amplitudes).

In the HO framework, there are various calculations of the transverse helicity amplitudes, among them we quote refs. \cite{cko,ki,cl,warns,sw}, while a systematic rHO approach has been used by \cite{ck}. A light cone calculation, using the CI \cite{ci} model, has been successfully applied to the $\Delta$ \cite{card_ND} and Roper excitation \cite{card_Roper}. For more recent light cone approaches, see ref. \cite{azn-bur} and references therein. The algebraic method, with the assumed charge distribution of the string, has been also used for the calculation of the transverse helicity amplitudes. The hCQM, in its non relativistic version, leads to nice predictions for the transverse excitation of the negative parity resonances \cite{aie2}. In both cases the theoretical curves exhibit a depletion at low $Q^2$, consistent with the results for the photocouplings.

A systematic description of the helicity amplitudes is still lacking. In this paper we present the {\bf predictions} of the non relativistic hCQM \cite{pl} for the longitudinal and transverse form factors for the excitation of fourteen baryon resonances, in comparison with the most recent experimental data. The curves for the transverse excitation of the negative parity resonances have been already published in \cite{aie2}, anyway we report them here again since in the last years new data have been published. The remaining curves are here published for the first time, although some of them have been shown in various conferences \cite{conf}, all of them without Proceedings.

\section{The hypercentral model}

We review briefly the hypercentral Constituent Quark Model, introduced in ref.
\cite{pl}, where the free parameters of the quark interaction have been fixed in
order to describe the non strange baryon spectrum. The same parameters have been used
for the calculation of various quantities (photocouplings \cite{aie}, transverse
helicity amplitudes for the excitation of negative parity resonances \cite{aie2} and elastic nucleon form factors \cite{mds2,rap}).These parameters are used for the present calculation of the longitudinal and transverse helicity amplitudes.

\begin{figure}[h]

\includegraphics[width=2.9in]{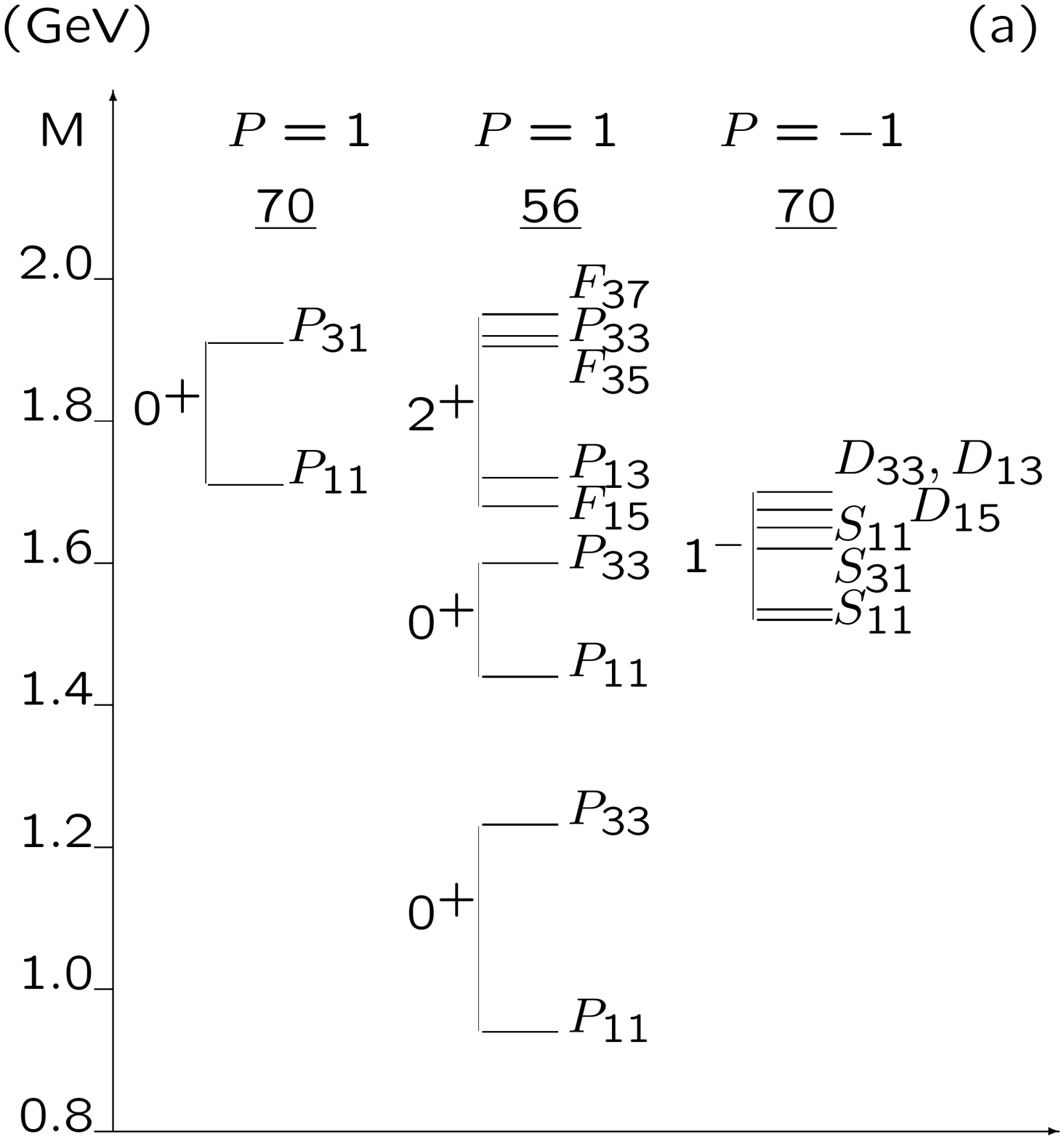}
\includegraphics[width=2.9in]{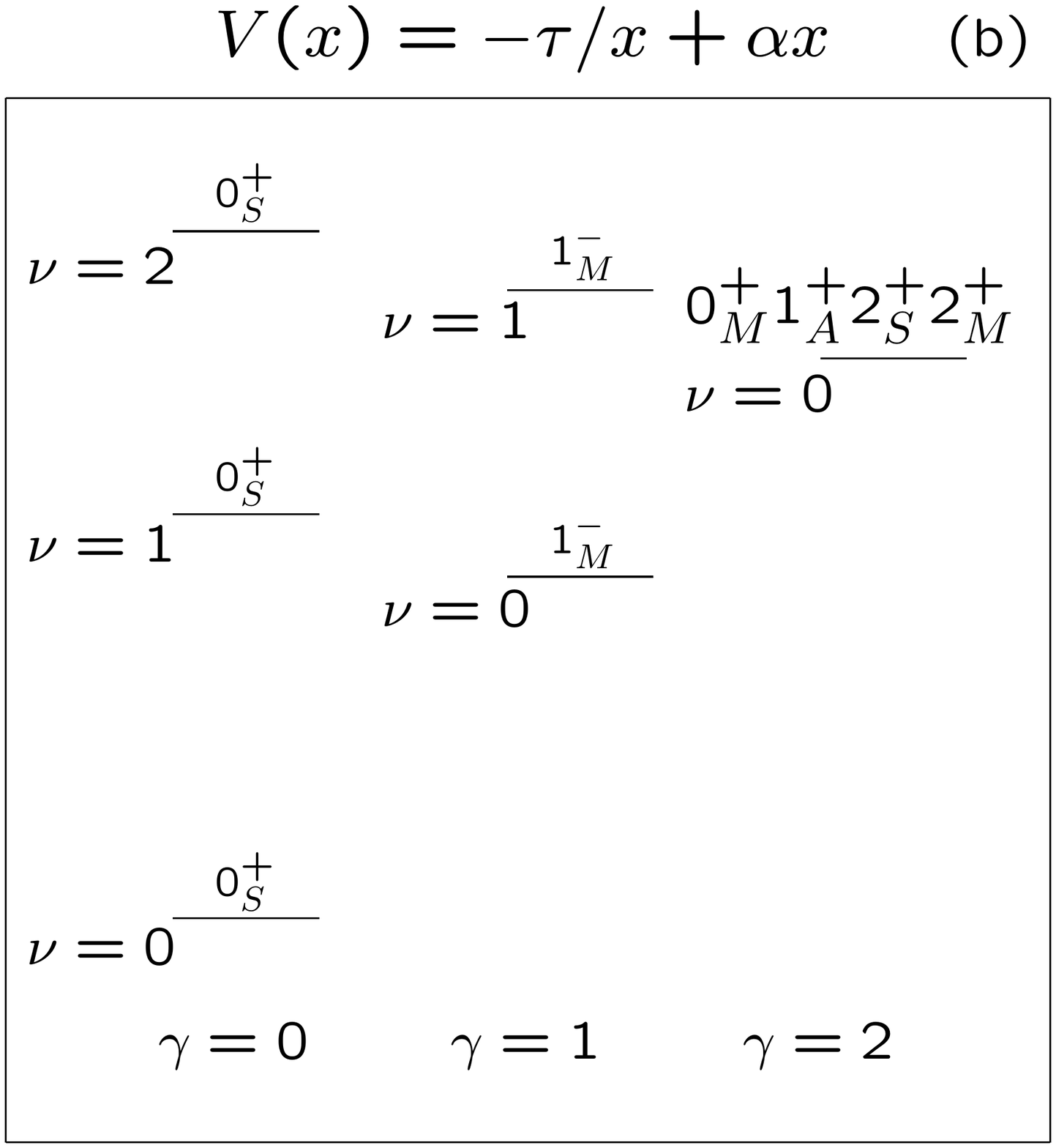}

\caption{ (Color online)(Left) The experimental spectrum of the non strange three- and four-star resonances \cite{pdg}. The states are reported in columns with the same parity P and grouped into $SU(6)$-multiplets. (Right) The spectrum given by the $SU(6)$ invariant part of  the hamiltonian Eq.(\ref{eq:ham}).}
\label{spect}
\end{figure}

 The $4-$ and $3-$star \cite{pdg} non strange resonances can be arranged in 
$SU(6)$ multiplets indicating that the quark dynamics has a
dominant $SU(6)$ invariant part accounting for the average multiplet
energies (see Fig.(\ref{spect}) left). The splittings within the multiplets are obtained by means of
a $SU(6)$ violating interaction, which can be spin and/or isospin
dependent and can be treated as a perturbation.

After removal of the center of mass coordinate, the space configurations of
three quarks in the non strange baryons
are described by the Jacobi coordinates, $\vec{\rho}$ and $\vec{\lambda}$,

\begin{equation}\label{eq:jac}
\vec{\rho}~=~ \frac{1}{\sqrt{2}}(\vec{r}_1 - \vec{r}_2) ~,\nonumber\\
~~~~\vec{\lambda}~=~\frac{1}{\sqrt{6}}(\vec{r}_1 + \vec{r}_2 - 2\vec{r}_3) ~. 
\end{equation}

One can introduce the hyperspherical coordinates, which are obtained by substituting
$\rho=|\vec{\rho}|$ and $\lambda=|\vec{\lambda}|$ with the hyperradius, $x$, and the
hyperangle,
$\xi$, defined respectively by 
\begin{equation}
x=\sqrt{{\vec{\rho}}^2+{\vec{\lambda}}^2} ~~, 
\xi=arctg(\frac{{\rho}}{{\lambda}}). 
\end{equation}
Using these coordinates, the kinetic term in
the three-body Schr\"{o}dinger equation can be rewritten as \cite{baf}
\begin{equation}
- \frac{1}{2m} ({\Delta}_{\rho}+{\Delta}_{\lambda})= - \frac{1}{2m}
(\frac{{\partial}^2}{{\partial}x^2}+\frac{5}{x} 
\frac{{\partial}}{{\partial}x}-\frac{L^2({\Omega}_{\rho},
{\Omega}_{\lambda},\xi)}{x^2})~~.~~\label{eq:ke}
\end{equation}
where $L^2(\Omega_{\rho},\Omega_{\lambda},\xi)$ is the six-dimensional generalization
of the squared angular momentum operator. Its eigenfunctions are the well known
hyperspherical harmonics
\cite{baf}
${Y}_{[{\gamma}]l_{\rho}l_{\lambda}}({\Omega}_{\rho},{\Omega}_{\lambda},\xi)$
 having eigenvalues $\gamma(\gamma+4)$, with
$\gamma=2n+l_{\rho}+l_{\lambda}$ (n is a non negative integer); they can be expressed
as products of standard spherical harmonics and Jacobi polynomials.

In the hypercentral Constituent Quark Model (hCQM) \cite{pl}, the $SU(6)$ invariant quark
interaction is assumed to depend on the hyperradius $x$ only $V_{inv}=V_{3q}(x)$. It has
been observed many years ago that a two-body quark-quark potential leads to matrix
elements in the baryon space quite similar to those of a hypercentral potential \cite{has}. On 
the other hand, a two body potential, treated in the hypercentral
approximation
\cite{hca}, that is averaged over angles and hyperangle, is transformed into a
potential which depends on $x$ only; in particular, a power-like two-body
potential
$\sum_{i<j}÷(r_{ij})^n$ in the hypercentral approximation is given by a term
proportional to
$x^n$. The hypercentral approximation has been shown to be valid, since it provides a
good description of baryon dynamics, specially for the lower states
\cite{hca}. 

The hyperradius
$x$ is a function of the coordinates of all the three quarks and then $V_{3q}(x)$
has also a three-body character. There are many reasons supporting the idea of
considering three-body interactions. First of all, three-body mechanisms are
certainly generated by the fundamental multi-gluon vertices predicted by QCD,
their explicit treatment is however not possible with the present
theoretical approaches and the presence of three-body mechanisms in
quark dynamics can be simply viewed as "QCD-inspired". Furthermore, flux tube models,
which have been proposed as a QCD-based description of quark interactions \cite{ip},
lead to Y-shaped three-quark configurations, besides the standard $\Delta-$like
two-body ones. A three-body confinement potential has  been shown to arise also if theÅ
quark dynamics is treated within a bag model \cite{hel}. Finally, it should be reminded
that threee-body forces have been considered also in the calculations by ref. \cite{ckp}
and in the relativized version of the Isgur-Karl model \cite{ci}.

For a hypercentral potential the three-quark wave function is factorized

\begin{equation}
\psi_{3q}(\vec{\rho},\vec{\lambda})~=
\psi_{\nu\gamma}(x)~~
{Y}_{[{\gamma}]l_{\rho}l_{\lambda}}({\Omega}_{\rho},{\Omega}_{\lambda},\xi)
\label{eq:psi}
\end{equation}
the hyperradial wave function is labeled by the grand angular quantum
number $\gamma$ defined above and  by the number of nodes $\nu$; it is obtained as 
 a solution of the hyperradial equation
\begin{equation}
[~\frac{{d}^2}{dx^2}+\frac{5}{x}~\frac{d}{dx}-\frac{\gamma(\gamma+4)}{x^2}]
~~{\psi}_{\nu\gamma}(x)
~~=~~-2m~[E-V_{3q}(x)]~~{\psi}_{\nu\gamma}(x)~~~.
\label{eq:rad}
\end{equation}

The Eq. (\ref{eq:rad}) can be solved analytically in two cases, the
first is the six-dimensional harmonic oscillator (h.o.)
\begin{equation}
\sum_{i<j}~\frac{1}{2}~k~(\vec{r_i} - \vec{r_j})^2~=~\frac{3}{2}~k~x^2~=
~V_{h.o}(x)
\label{eq:ho}
\end{equation}
and the hyperCoulomb (hC) potential 
\begin{equation}
V_{hyc}(x)= -\frac{\tau}{x}. 
\label{eq:hyc}
\end{equation}

The hypercoulomb term $1/x$ has important
features \cite{pl,sig}. First of all, the negative parity states are exactly
degenerate with the first positive parity excitations. The observed Roper resonance is
somewhat lower with respect to the negative parity baryon resonance, at variance with
the prediction of any $SU(6)-$invariant two-body potential, therefore the
hypercoulomb potential provides a good
starting point for the description of the spectrum. Moreover, the
resulting form factors have a power-law behaviour, again leading to an improvement 
with respect to the widely used harmonic oscillator. 

The hypercentral model potential includes a confinement term which is
linear in $x$
\begin{equation}\label{eq:pot}
V(x)= -\frac{\tau}{x}~+~\alpha x.
\end{equation}

 Interactions of the type
linear plus Coulomb-like have been used for the meson sector, e.g. the
Cornell potential and have been supported recently by  Lattice QCD calculations
\cite{bali}. In this respect, the potential Eq. (\ref{eq:pot}) can be considered as 
the hypercentral approximation of the lattice QCD potential.

The splittings within the multiplets are produced by a
perturbative term breaking
$SU(6)$, which can be assumed to be the standard hyperfine interaction $H_{hyp}$
\cite{ik}.
The three quark hamiltonian is then:
\begin{equation}\label{eq:ham}
H = \frac{p_{\lambda}^2}{2m}+\frac{p_{\rho}^2}{2m}-\frac{\tau}{x}~+~\alpha x+H_{hyp},
\end{equation}
where $m$ is the quark mass (taken equal to $1/3$ of the nucleon mass). The strength
of the hyperfine interaction is determined in order to reproduce the $\Delta-N$ mass
difference, the remaining two free parameters are fitted to the spectrum, leading to
the following values \cite{pl}:
\begin{equation}\label{eq:par}
\alpha= 1.16fm^{-2},~~~~\tau=4.59~.
\end{equation}
\noindent the spectrum given by the $SU(6)$ invariant part of the hamiltoniana Eq. (\ref{eq:ham}) is reported in the right part of Fig.(\ref{spect}). The degeneracy between the $0_S^+$ and $1_M^-$ states,  typical of the hypercoulomb interaction, is removed by the confinement term.

Having fixed the parameters of the potential, the hyperradial wave functions for the
ground and excited states can be calculated and therefore one can build up the
three-quark states for the various resonances, taking due account of the antisymmetry
requirements. In fact, the complete three-quark wave function can in general be
factorized in four parts, that is the colour, spin, flavour (isospin for non strange
baryons) and space factors:
\begin{equation}\label{eq:psi_tot} 
\Psi_{3q}÷=÷\theta_{colour}÷÷÷ \cdot \chi_{spin}÷÷÷ \cdot \Phi_{isospin}÷÷÷
   \cdot \psi_{3q}(\vec{\rho},\vec{\lambda}).
\end{equation}
The colour wave function must be completely antisymmetric in order to give rise to
colourless baryons, therefore the remaining factors have to be combined to an overall
symmetric function. 

The introduction of $SU(6)-$configurations is beneficial, since
in this way the spin and flavour states are combined to form a unique $SU(6)-$state,
which must share the same symmetry property with the space wave
function. In Appendix A we give
the explicit form of the $SU(6)-$configurations describing the various baryon
states. Since the h.o. is a hypercentral potential as well, the
$SU(6)-$configurations are obtained by means of the same procedure followed in ref.
\cite{ik}, simply substituting ${\psi}_{\nu\gamma}^{h.o.}(x)$ with  ${\psi}_{\nu\gamma}(x)$, the eigenfunctions 
obtained with the hypercentral potential of eq.(\ref{eq:pot}). 

The mixing is provided by the hyperfine interaction in Eq. (\ref{eq:ham})
and the mixing coefficients are obtained fitting the observed baryon spectrum.
The model is in this way completely fixed and the knowledge of the states of all
resonances allows a systematic calculations of various physical quantities of
interest. This systematic analysis has already been performed for the phocouplings
\cite{aie}, the transverse electromagnetic transition amplitudes  \cite{aie2}, the
elastic nucleon form factors
\cite{mds} and the ratio between the electric and magnetic proton form factors
\cite{rap}. As far as the transverse electromagnetic form factors, in ref.
\cite{aie2} only the negative parity states have been considered, however we have
results for the transition to all resonances.

The three-quark interaction of eq. (\ref{eq:pot}) has been recently used in fully relativistic approach in point from in order to describe the elastic electromagnetic form factors of the nucleon \cite{ff_07,ff_10}

 In this paper we present the parameter
free calculation of the longitudinal and transverse transition form factors using the potential of
Eq. (\ref{eq:pot}) with the parameters reported in Eq. (\ref{eq:par}).

\section{The helicity amplitudes}

The electromagnetic transition amplitudes, 
$A_{1/2}$, $A_{3/2}$ and $S_{1/2}$, are defined as the matrix elements of
the quark electromagnetic interaction, $A_{\mu} ÷J^{\mu}$, between the nucleon, 
$N$, and the resonance, $B$, states:
\begin{equation} \label{eq:hel}
\begin{array}{rcl}
\mathcal{A}_{1/2}&=&   \sqrt{\frac{2 \pi \alpha}{k}}   \langle B, J', J'_{z}=\frac{1}{2}\ | J_{+}| N, J~=~
\frac{1}{2}, J_{z}= -\frac{1}{2}\
\rangle\\
& & \\
\mathcal{A}_{3/2}&=& \sqrt{\frac{2 \pi \alpha}{k}}  \langle B, J', J'_{z}=\frac{3}{2}\ |J_{+} | N, J~=~
\frac{1}{2}, J_{z}= \frac{1}{2}\
\rangle\\
& & \\
\mathcal{S}_{1/2}&=&   \sqrt{\frac{2 \pi \alpha}{k}}   \langle B, J', J'_{z}=\frac{1}{2}\ | J_{0}| N, J~=~
\frac{1}{2}, J_{z}= \frac{1}{2}\
\rangle\\\end{array}
\end{equation}
$J_{\mu}$ is the electromagnetic current carried by quarks and will be used in its non relativistic  form 
\cite{cko, ki}; $k$ is the virtual photon momentum in the Breit frame. For the transverse excitation, the photon has been assumed,  without loss of generality, as left-handed. Moreover the z-axis is assumed along the virtual photon momentum.

In order to compare the theoretical calculations with data, one has to consider that the helicity amplitudes extracted from the meson photoproduction contain also the sign of the $\pi N N^*$ vertex. The theoretical helicity amplitudes are therefore defined up to a common phase factor $\zeta$
\begin{equation}
A_{1/2,3/2}~=~\zeta~ \mathcal{A}_{1/2,3/2} ~~~~~~~~~~~S_{1/2}~=~\zeta~ \mathcal{S}_{1/2} 
\end{equation}
The factor $\zeta$ will be taken in agreement with the choice of ref. \cite{ki}, with the exception of the Roper resonance, in which case the sign is in agreement with the analysis performed in \cite{azn07}.

In the following we report the results of the calculations for those resonances which, according to the PDG classification \cite{pdg}, have an electromagnetic decay with a three- or four- star status. This happens for twelve resonances, namely the $I=\frac{1}{2}$ states 
\begin{equation}
P11(1440), ÷D13(1520),÷S11(1535), ÷S11(1650), ÷D15(1675), ÷F15(1680), ÷P11(1710)
\end{equation}
 and the $I=\frac{3}{2}$ ones 
 \begin{equation}
 P33(1232),S31(1620), D33(1700), F35(19005), F37(1950)
 \end{equation}
 Besides these states, we  have considered also the states $D13(1700)$ and $P13(1720)$, which are excited in an energy range particularly interesting for the phenomenological analysis.

The calculations of the matrix elements of Eqs.(\ref{eq:hel}) are performed using as baryon states the eigenstates of the hamiltonian (\ref{eq:ham}). For each resonance, in Appendix A we list the states in the $SU(6)$ limit; the physical states of the various resonances are given by the configuration mixing produced by the hyperfine interaction in Eq.(\ref{eq:ham}).

It should be stressed that, after having fixed the free parameters  (see Eq. (\ref{eq:par}) in order to reproduce the baryon spectrum, the baryon states are completely determined and the results for the helicity amplitudes reported in the following sections are parameter free predictions of the hypercentral Constituent Quark Model.

\subsection{The photocouplings}

The proton and neutron photocouplings predicted by the hCQM \cite{aie} are reported in Tables 1,2 and 3 in comparison with the PDG data \cite{pdg} . The overall behaviour is fairly well reproduced, but in general there is a lack of strength. The proton transitions to the S11(1650), D15(1675) and D13(1700) resonances vanish exactly in absence of hyperfine mixing and are therefore entirely due to the SU(6) violation. The results obtained with other calculations are qualitatively not much different \cite{aie,cr2} and this is because the various CQM models have the same SU(6) structure in common.

\begin{table}
\centering
\caption[]{Photocouplings (in units $10^{-3} GeV^{-1/2}$) predicted by the hCMQ in comparison with PDG data for proton excitation to N*-like resonances. The proton transitions to the S11(1650), D15(1675) and D13(1700) resonances vanish in the SU(6) limit.}
\vspace{15pt}
\label{photo}
\begin{tabular}{|c|rr|rr|}
\hline
& & & &   \\
$Resonance$ &$A_{1/2}^p (hCQM)$ & $A_{1/2}^p (PDG)$ & $A_{3/2}^p (hCQM)$ & $A_{3/2}^p (PDG)$ \\
& & &  &  \\
\hline
& & &  & \\
P11(1440)&$88$ & $-65 \pm 4$  & &  \\
D13(1520)&$-66$ & $-24 \pm 9$ & $67$ & $ 166 \pm 5$  \\
S11(1535)&$109$ & $90 \pm 30$ &  & \\
S11(1650)&$69$ & $53 \pm 16$ &  & \\
D15(1675)&$1$ & $19 \pm 8$ & $2$ &$15 \pm 9$   \\
F15(1680)&$-35$ & $ -15 \pm 6$ & $24$ & $ 133 \pm 12$  \\
D13(1700)&$8$ & $ -18 \pm 13$ & $-11$ &$ -2  \pm 24$  \\
P11(1710)&$43$ &$ 9 \pm 22$ & & \\
P13(1720)&$94$ & $ 18 \pm 30 $ &$-17$ & $ -19 \pm 20$ \\
& & & & \\
\hline
\end{tabular}
\end{table}

\begin{table}
\centering
\caption[]{ The same as Table 1 for neutron excitation}
\vspace{15pt}
\label{photo}
\begin{tabular}{|c|rr|rr|}
\hline
& & & &   \\
$Resonance$ &$A_{1/2}^n (hCQM)$ & $A_{1/2}^n (PDG)$ & $A_{3/2}^n (hCQM)$ & $A_{3/2}^n (PDG)$ \\
& & &  &  \\
\hline
& & &  & \\
P11(1440)&$58$ & $40 \pm 10$  & &  \\
D13(1520)&$-1$ & $-59 \pm 9$ & $-61$ & $ -139 \pm 11$  \\
S11(1535)&$-82$ & $-46 \pm 27$ &  & \\
S11(1650)&$-21$ & $-15 \pm 21$ &  & \\
D15(1675)&$-37$ & $-43 \pm 12$ & $-51$ &$-58 \pm 13$   \\
F15(1680)&$38$ & $ 29 \pm 10$ & $15$ & $ -33 \pm 9$  \\
D13(1700)&$12$ & $ 0 \pm 50$ & $70$ &$ -3  \pm 44$  \\
P11(1710)&$-22$ &$ -2 \pm 14$ & & \\
P13(1720)&$-48$ & $ 1 \pm 15 $ &$4$ & $ -29 \pm 61$ \\
& & & & \\
\hline
\end{tabular}
\end{table}

\begin{table}
\centering
\caption[]{The same as Table 1 for the excitation to $\Delta$-like resonances}
\vspace{15pt}
\label{photo}
\begin{tabular}{|c|rr|rr|}
\hline
& & & &  \\
$Resonance$ &$A_{1/2}^p (hCQM)$ & $A_{1/2}^p (PDG)$ & $A_{3/2}^p (hCQM)$ & $A_{3/2}^p (PDG)$ \\
& & &  & \\
\hline
& & &  & \\
P33(1232)&$-97$ & $-135 \pm 6$ & $-169$ &$-250 \pm 8$  \\
S31(1620)&$30$ &$27 \pm 11$ &  & \\
D33(1700)&$81$ & $104 \pm 5$ &$70$ & $85 \pm 2$ \\
F35(1905)&$-17$ & $26 \pm 11$ &$-51$ & $-45 \pm 20$ \\
F37(1950)&$-28$ & $-76 \pm 12$ & $-35$ & $-97 \pm 10$ \\
& & & &  \\
\hline
\end{tabular}
\end{table}

\subsection{The transition form factors}
Taking into account the $Q^2-$behaviour of the transition matrix elements, one can
calculate the hCQM helicity amplitudes \cite{aie2}. 

In order to compare with the experimental data, the calculation should be performed in the rest frame of the resonance (see e.g. \cite{azn-bur-12}). The nucleon and resonance wave functions are calculated in their respective rest frames and, before evaluating the matrix elements given in Eqs. (\ref{eq:hel}), one should boost the nucleon to the resonance c.m.s.. In our non relativistic approach such boost is trivial but not correct, because of the large nucleon recoil. In order to minimize the discrepancy between the non relativistic and the relativistic boost in comparing with the experimental data, we use the Breit frame, as in refs. (\cite{aie2,bil}). Therefore we use the following kinematic relation:
\begin{equation}
{\vec{k}}^2 = Q^2 + \frac{(W^2 - M^2)^2}{2(M^2 + W^2) + Q^2}~,
\label{eq:breit}
\end{equation}
where $M$ is the nucleon mass, $W$ is the mass of the resonance, $k_0$ and $\vec{k}$ are the energy and the momentum of the virtual photon, respectively, and $Q^2~=
~{\vec{k}}^2- k_0^2$. For consistency reasons, in the calculations  we have used the value of $W$ given by the model and not the phenomenological ones.

The matrix elements of the e.m. transition operator between 
any two 3q-states are expressed in terms of integrals involving the 
hyperradial wavefunctions and are calculated numerically. The computer code 
has been tested by comparison with the analytical results obtained with the 
h.o. model of Refs. \cite{cko, ki} and with the analytical model of Ref. 
\cite{sig}.

\subsection{The excitation to the $\Delta$ resonance}

\begin{figure}[h]

\includegraphics[width=3in]{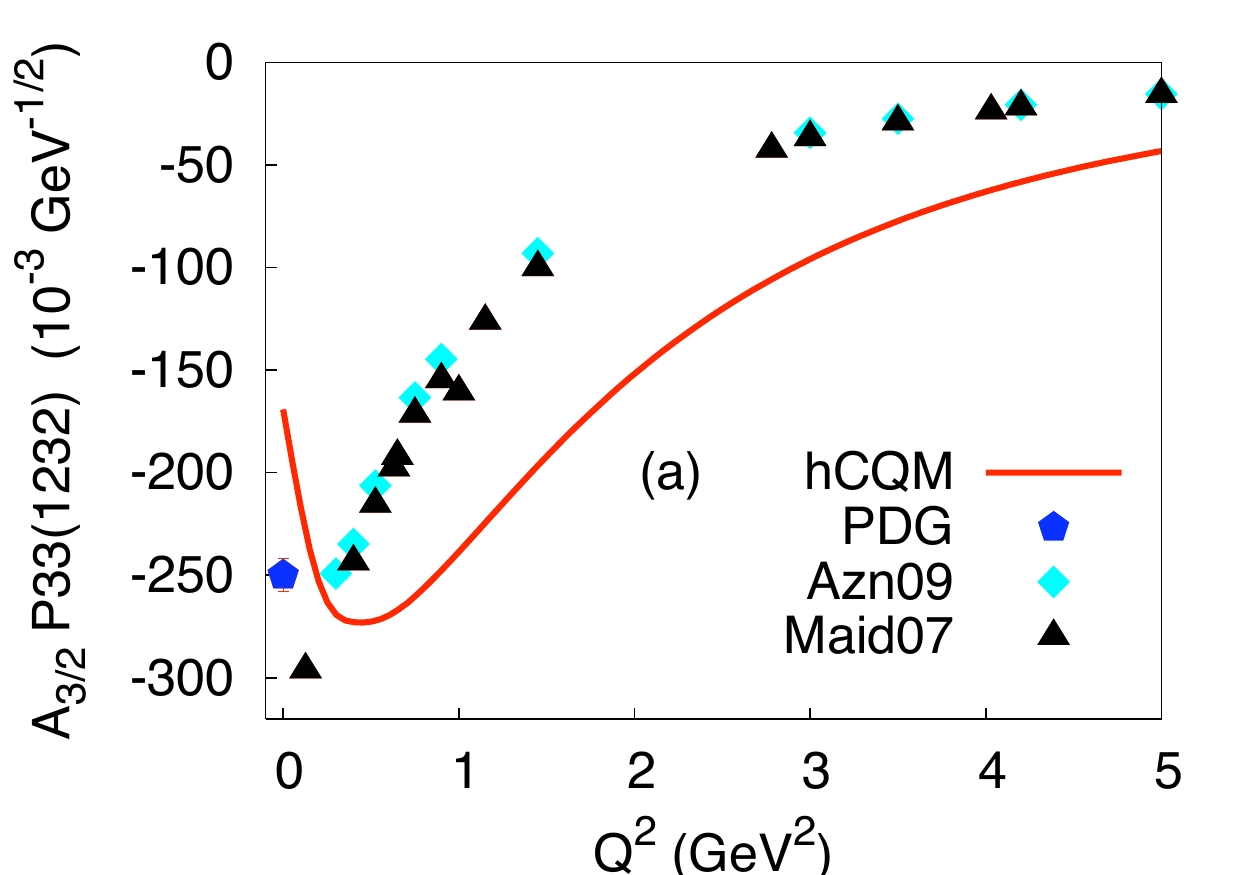}
\includegraphics[width=3in]{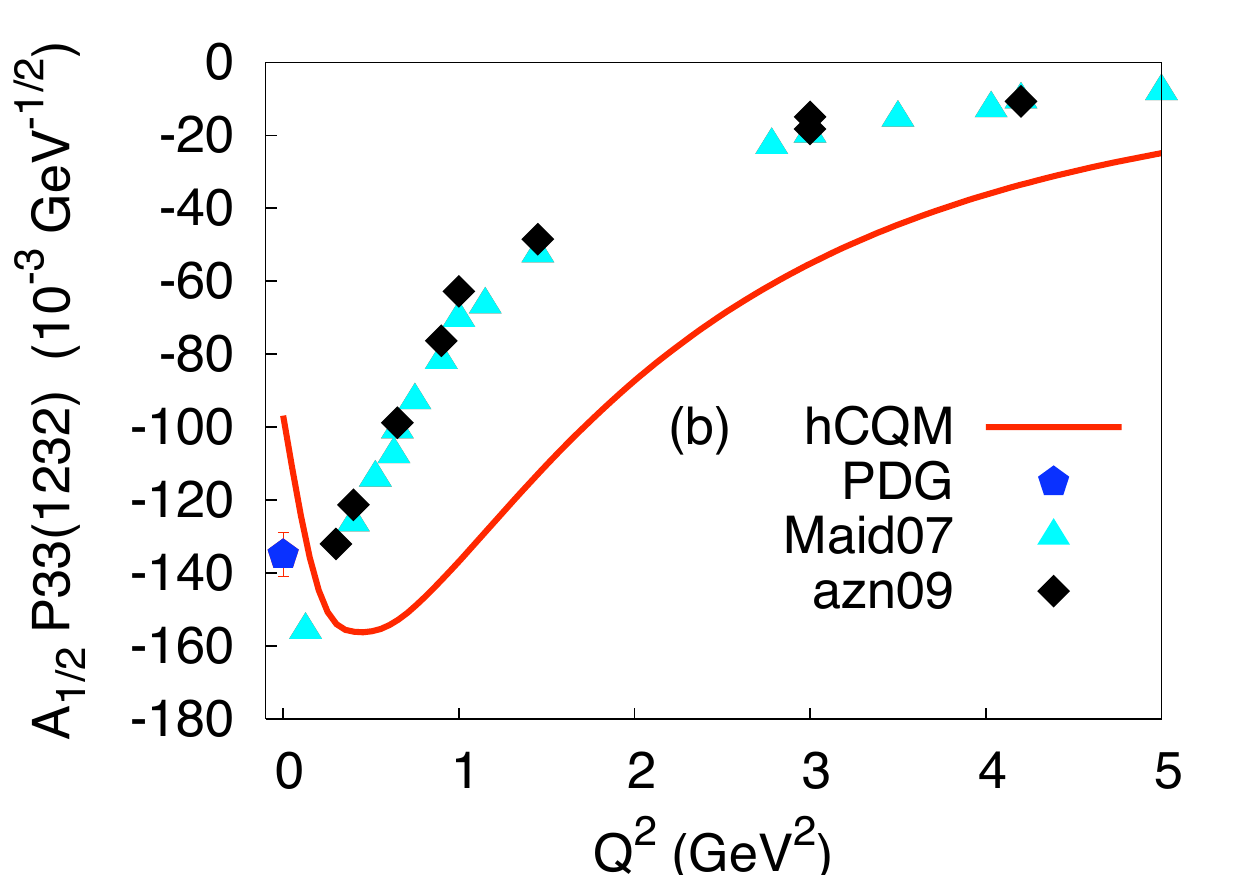}
\includegraphics[width=3in]{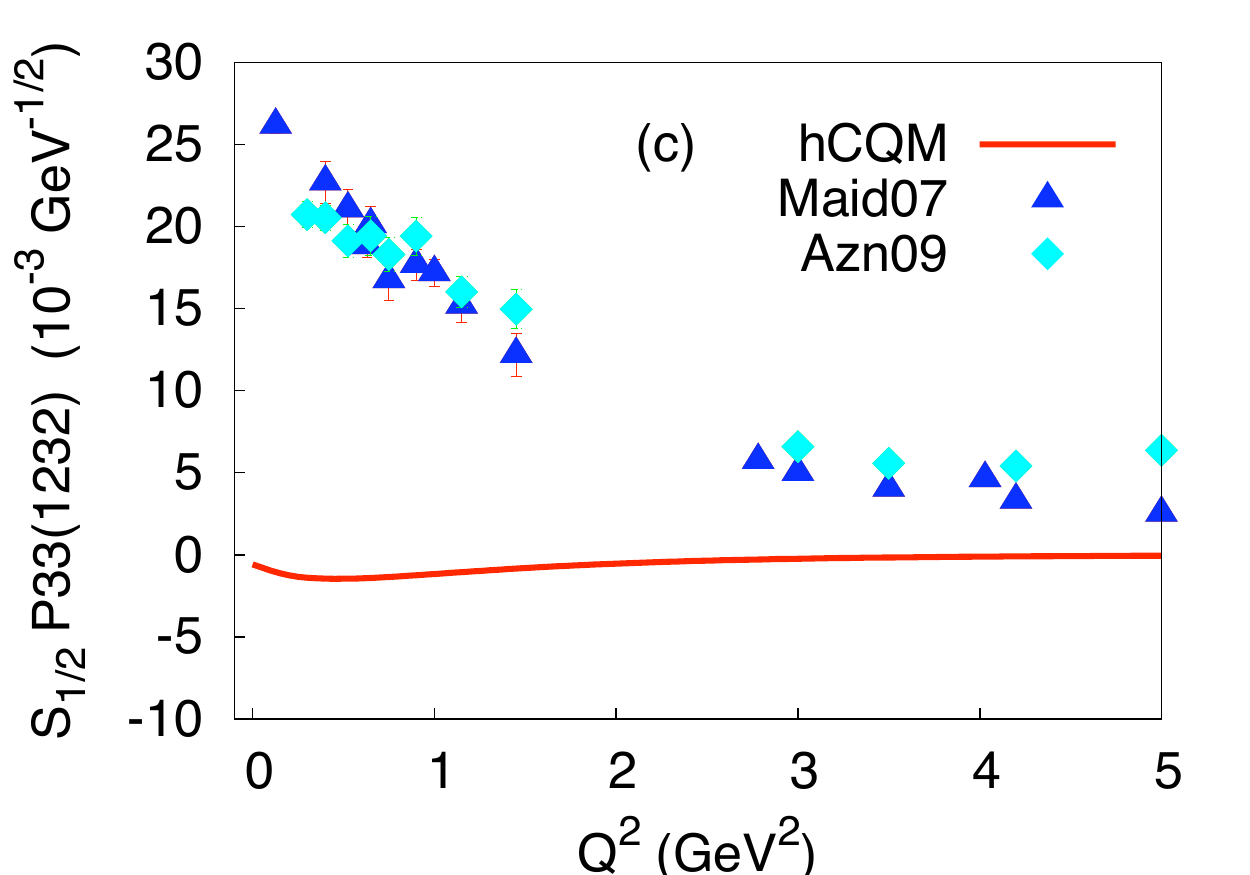}

\caption{(Color on line)The P33(1232) helicity amplitudes predicted by the hCQM (full curves)  $A_{3/2}$ (a), $A_{1/2}$ (b) and $S_{1/2}$ (c), in comparison with the data of ref. \cite{azn09} and with the the Maid2007 analysis \cite{maid07} of the data by refs. \cite{joo02} and \cite{lav04}. The PDG points \cite{pdg} are also shown.}
\label{p33}
\end{figure}

The $N-\Delta$ helicity amplitudes are shown in Fig. \ref{p33}.
The transverse  excitation to the $\Delta$ resonance has a lack of strength at low $Q^2$, a feature in common with all CQM calculations. The medium-high  $Q^2$ behavior is decreasing too slowly with respect to data, similarly to what happens for the nucleon elastic form factors \cite{mds,ff_07}. In this case, the non relativistic calculations are improved by taking into account relativistic effects. Since the $\Delta$ resonance  and the nucleon are in the ground state $SU(6)$-configuration, we expect that their internal structures  have strong similarities and that a good description of the $N-\Delta$ transition from factors is possible only with a relativistic approach. Such feature is further supported by the fact that the transitions to the higher resonances are only slightly affected by relativistic 
effects \cite{mds}.

An important issue in connection with the $\Delta$ resonance is the possible deformation, which manifests itself in a non zero value for the transverse and longitudinal quadrupole strength. To this end one considers in particular the ratio
\begin{equation} \label{eq:REM}
R_{EM}~=~- ~\frac{G_{E}}{G_{M}}~=~- ~\frac{\sqrt{3} ~ A_{1/2}~-~A_{3/2}}{\sqrt{3}~A_{1/2}~+~A_{3/2}}
\end{equation}
\noindent where $G_{E}$ and $G_{M}$ are, respectively, the transverse electric and magnetic form factors for the $N \rightarrow \Delta$ transition \cite{dev}. If the quarks in the nucleon and the
$\Delta$ are in a pure $S-$wave state there is no quadrupole
excitation \cite{bm}. A deformation can be produced if the 
interaction contains a hyperfine term as in Eq. (\ref{eq:ham}) and both the nucleon and the $\Delta$ states acquire D-components..

At the photon point, the experimental value of the ratio is $R_{EM}~=~-0.025 \pm 0.005$ \cite{pdg}, which is not far from the value given by  \cite{ki, ikk}. It should be mentioned that,  taking into account  the higher shells and the 
Siegert's theorem for a more accurate and reliable calculation, the value
$R=0.02$ was obtained \cite{dg}.

In our model the ratio Eq.(\ref{eq:REM}) is about $0.005$, which means that the  deformation is very low. This fact is confirmed by the small theoretical value of the longitudinal quadrupole transition amplitude $S_{1/2}$ (see Fig. \ref{p33} (bottom)). As stated above, the introduction of relativity is expected to be beneficial, but  the discrepancy may be due to a quite different reason. 

An alternative approach to baryon resonance physics is provided by dynamical models (see e.g. \cite{dmt,sato} and references therein). The calculations performed with the DMT model \cite{dmt} have shown that  the  $N-\Delta$ $S_{1/2}$ transition amplitude is almost completely determined by the  pionic meson cloud \cite{ts03}. Actually also for many other transitions the meson cloud  seems to give important contributions in correspondence of  lack of strength of the hCQM \cite{ts03}. This leads to the problem of missing degrees of freedom in the CQM calculations, but we shall come back on this topic later on.

\subsection{The proton excitation to the second resonance region}

\subsubsection{The Roper}

\begin{figure}[h]

\includegraphics[width=3in]{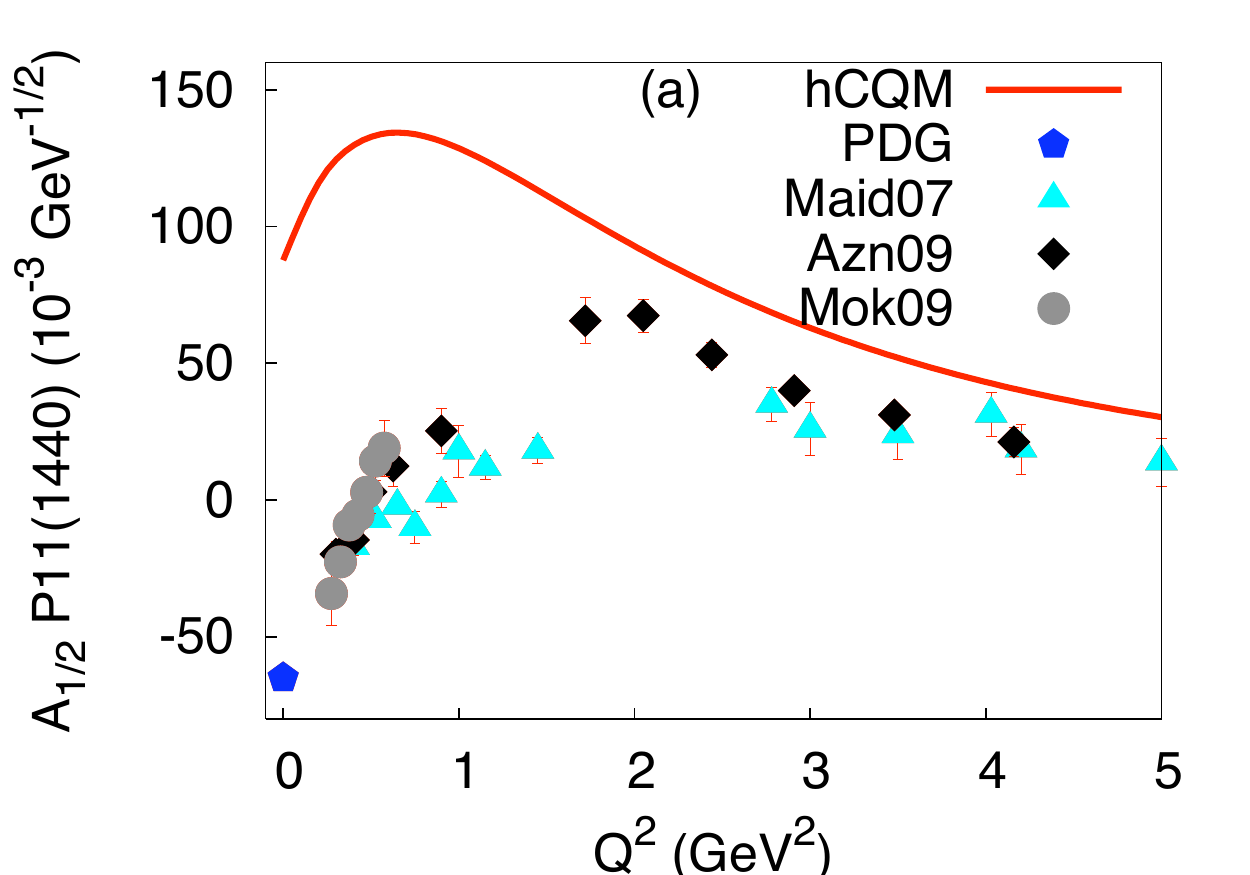}
\includegraphics[width=3in]{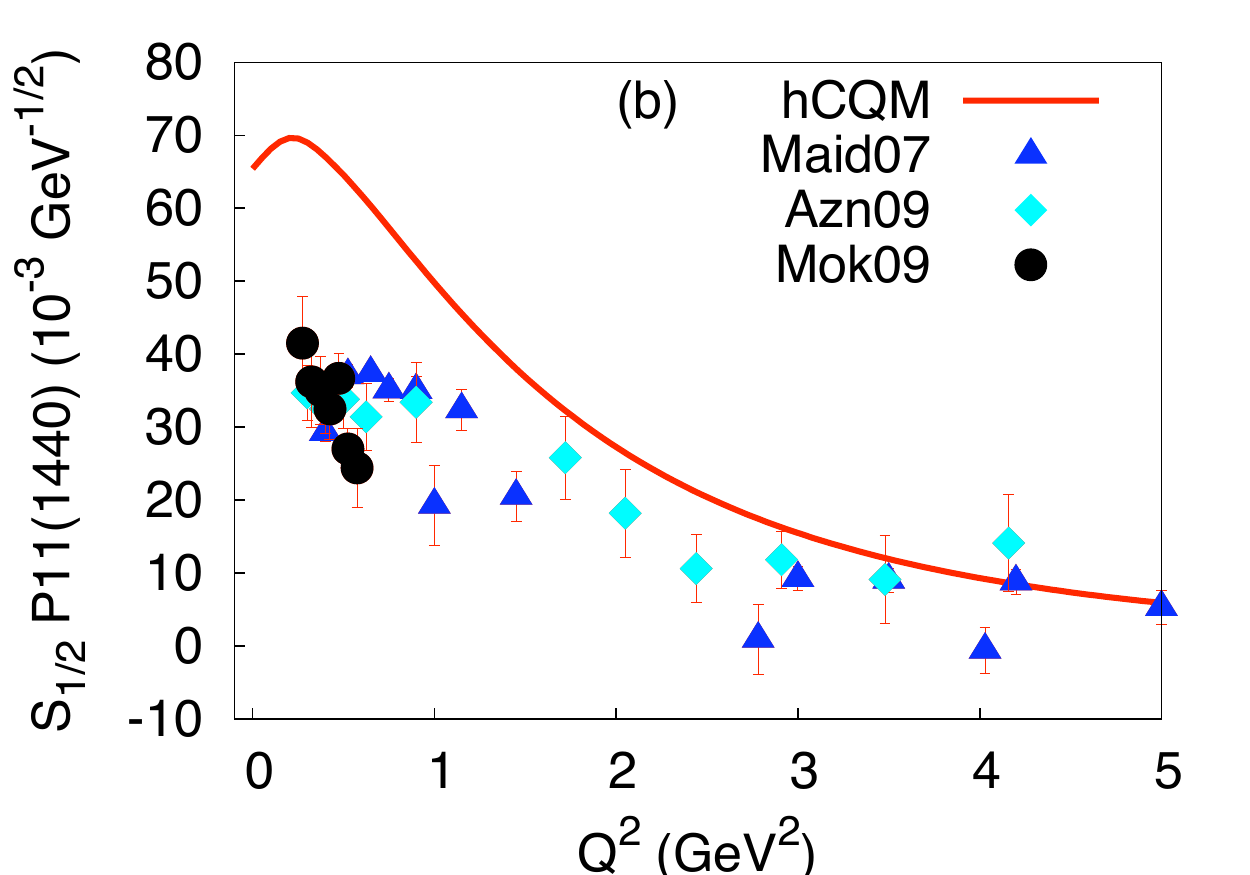}

\caption{(Color on line)The P11(1440) proton transverse (a) and longitudinal (b) helicity amplitudes predicted by the hCQM (full curves), in comparison with the data of refs.  \cite{vm09}, \cite{azn09} and the Maid2007 analysis \cite{maid07}  of the data by refs. \cite{fro99},\cite{joo02},  \cite{lav04} and \cite{ung06}. The PDG point \cite{pdg} is also shown.}
\label{p11}
\end{figure}

Because of the $\frac{1}{x}$ term in the hypercentral potential of eq.(\ref{eq:ham}), the Roper resonance can be accomodated in the first resonance region, at variance with h.o. models, which predict it to be a 2 $\hbar \omega$ state. The results for the helicity amplitudes are shown in Fig.(\ref{p11}). There are problems in the low $Q^2$ region, but for the rest the agreement is interesting, specially if one remember that the curves are predictions and the Roper has been often been considered a crucial state, non easily included into a constituent quark model description. In the past, exotic explanations of the Roper have been introduced, in particular a model of the Roper as a three-quark-gluon structure has been proposed \cite{li-bu}; however such model predicts a vanishing value for the longitudinal excitation \cite{azn08}, a result which is ruled out by the data shown in Fig.(\ref{p11}). In the present model, the Roper is a hyperradial excitation of the nucleon.

\subsubsection{The negative parity states}

It should be stressed that, apart from the case of the $D13(1700)$ and $D15(1675)$ resonances, the transverse helicity curves are the same as in ref. \cite{aie2}.

\begin{figure}[h]

\includegraphics[width=3in]{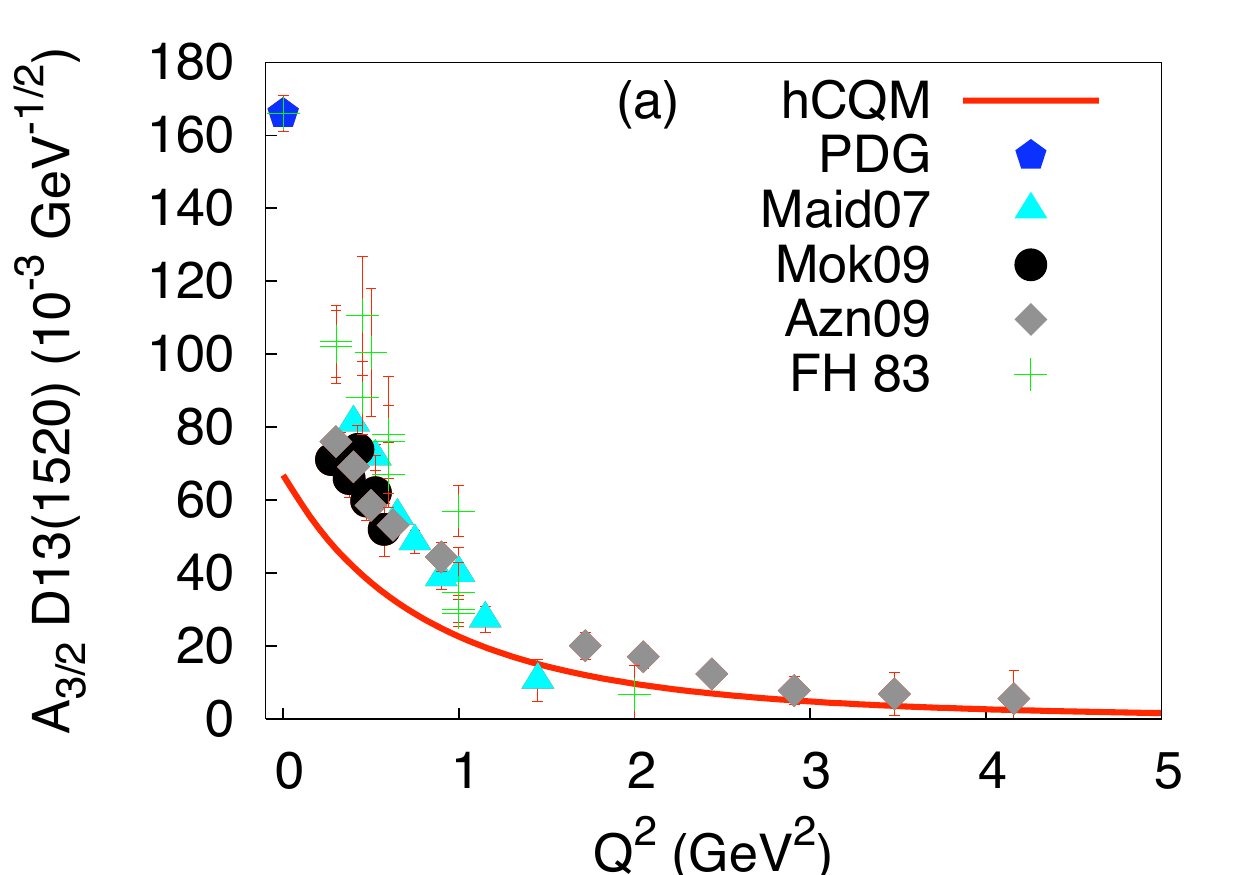} 
\includegraphics[width=3in]{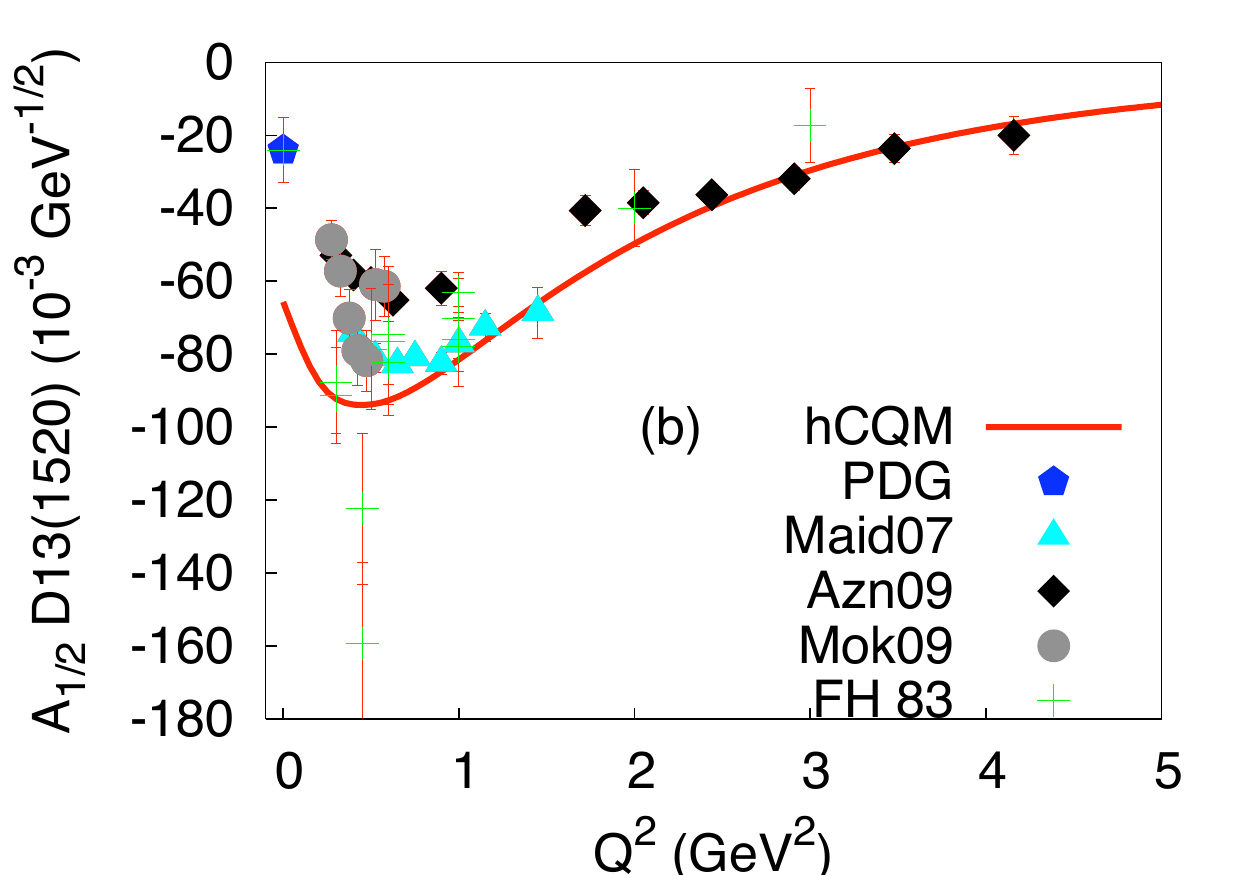}
\includegraphics[width=3in]{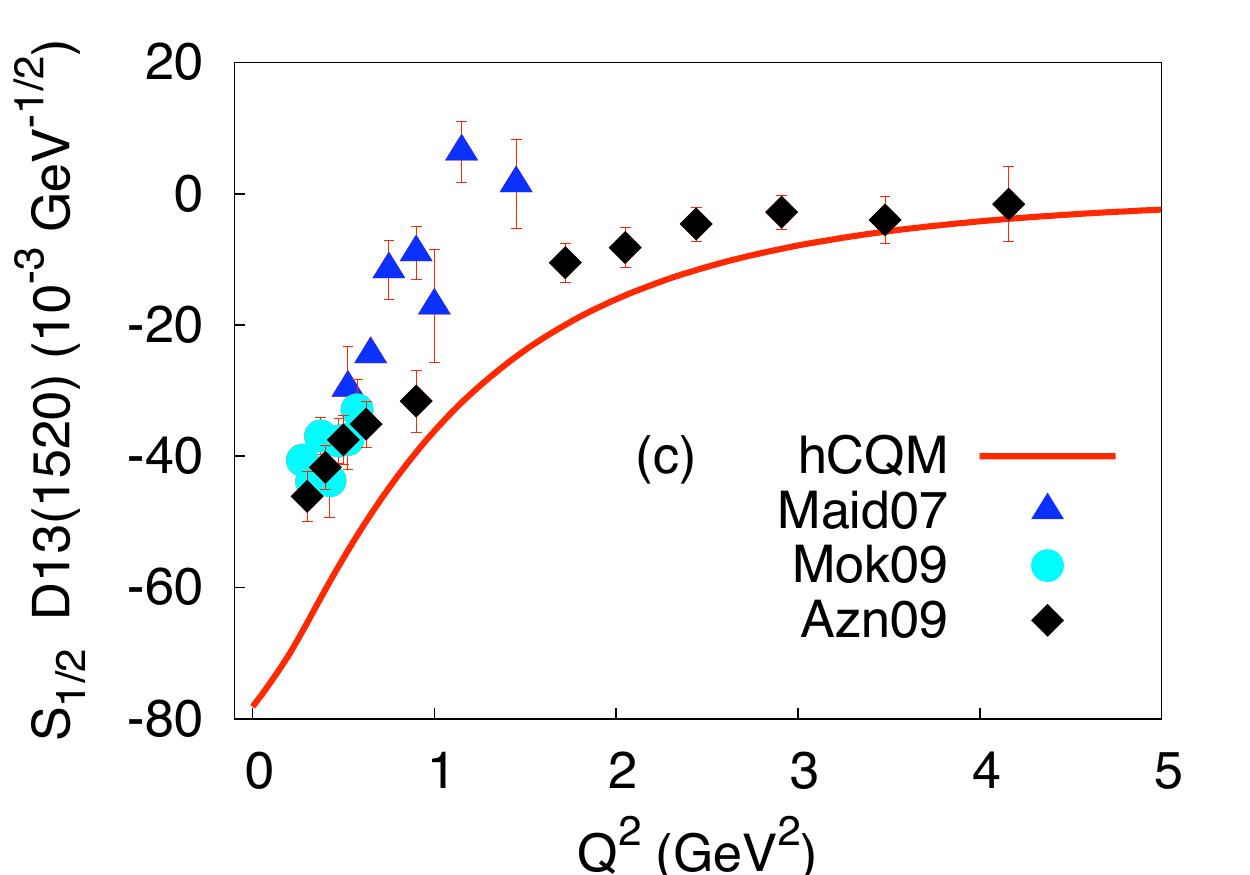}

\caption{(Color on line)The D13(1520)  proton helicity amplitudes predicted by the hCQM (full curves)  $A_{3/2}$ (a), $A_{1/2}$ (b) and $S_{1/2}$ (c), in comparison with the data of refs. \cite{vm09}, \cite{azn09}, with the compilation reported in refs. \cite{fh,ger} and the Maid2007 analysis \cite{maid07} of the data by refs. \cite{joo02} and \cite{lav04}. The PDG points \cite{pdg}are also shown.}
\label{d13}
\end{figure}

\begin{figure}[h]

\includegraphics[width=3in]{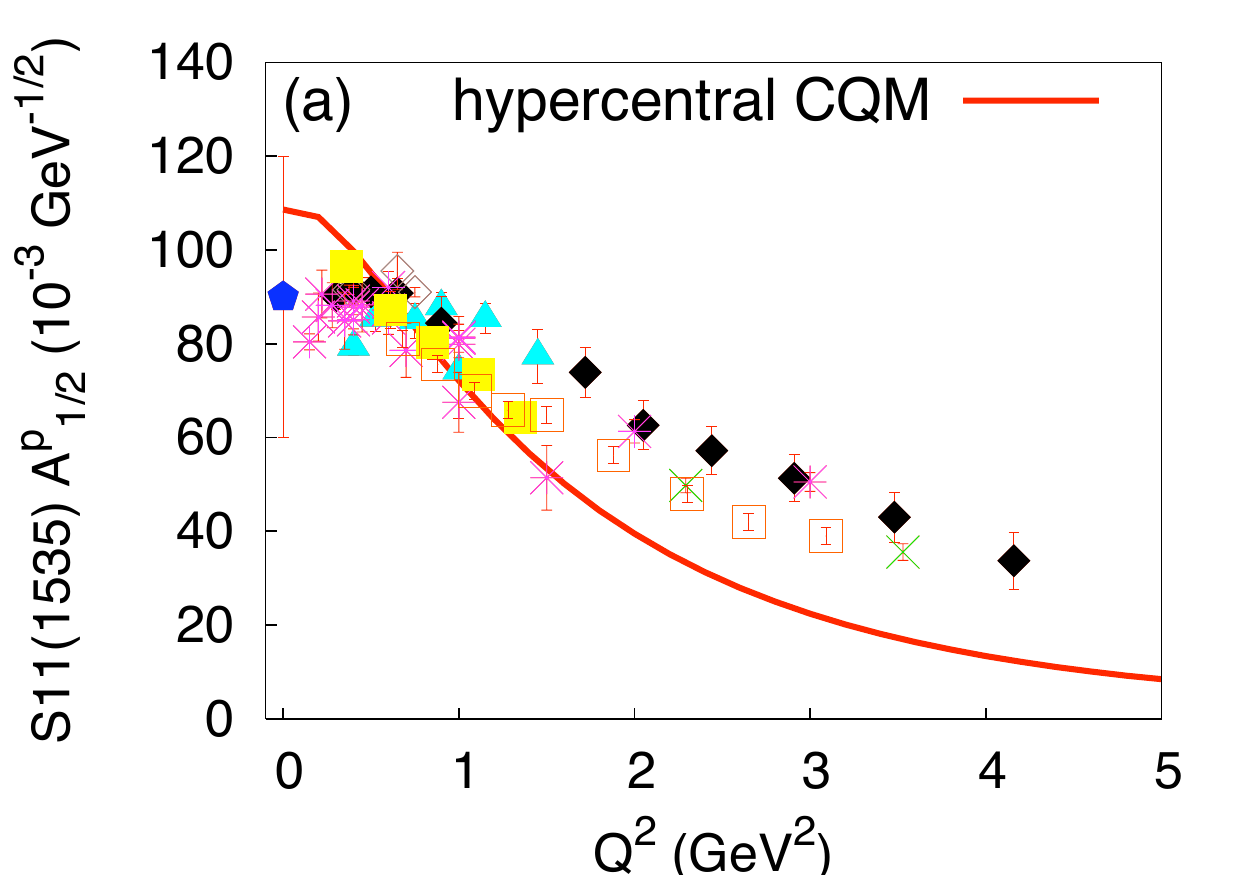}
\includegraphics[width=3in]{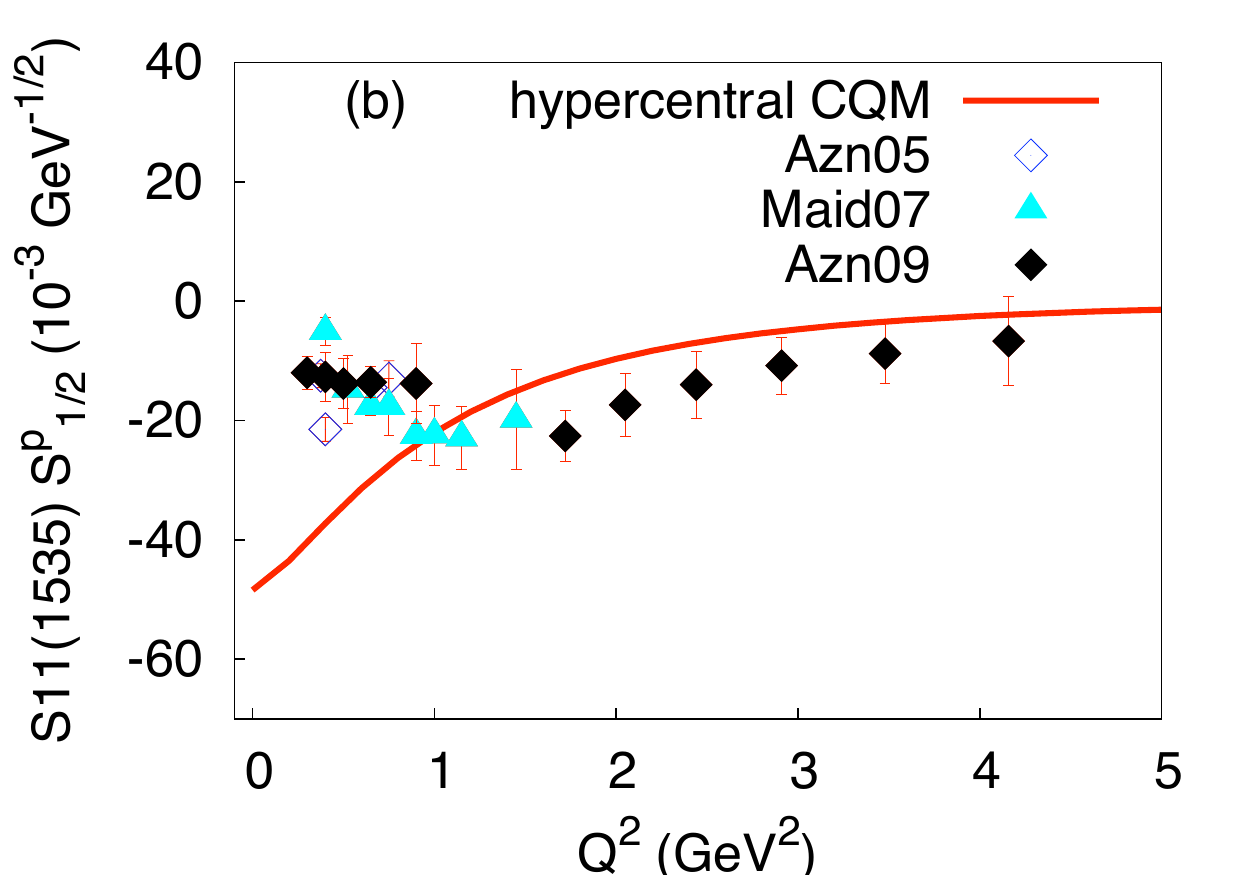}
\caption{(Color on line)The S11(1525) proton transverse (a) and longitudinal (b) helicity amplitudes predicted by the hCQM (full curve), in comparison with the data of refs. \cite{azn05_1} (open diamonds),  \cite{azn09} (full diamonds), \cite{arm99} (crosses), \cite{den07} (open squares), \cite{thom01} (full squares), the Maid2007 analysis \cite{maid07} (full triangles) of the data by refs. \cite{joo02} and the compilation of the Bonn-Mainz-DESY data of refs. \cite{kru,bra,beck,breu} (stars), presented in \cite{thom01}. The PDG point \cite{pdg} (pentagon) is also shown.}
\label{s11}

\end{figure}

\begin{figure}[]
 
\includegraphics[width=3in]{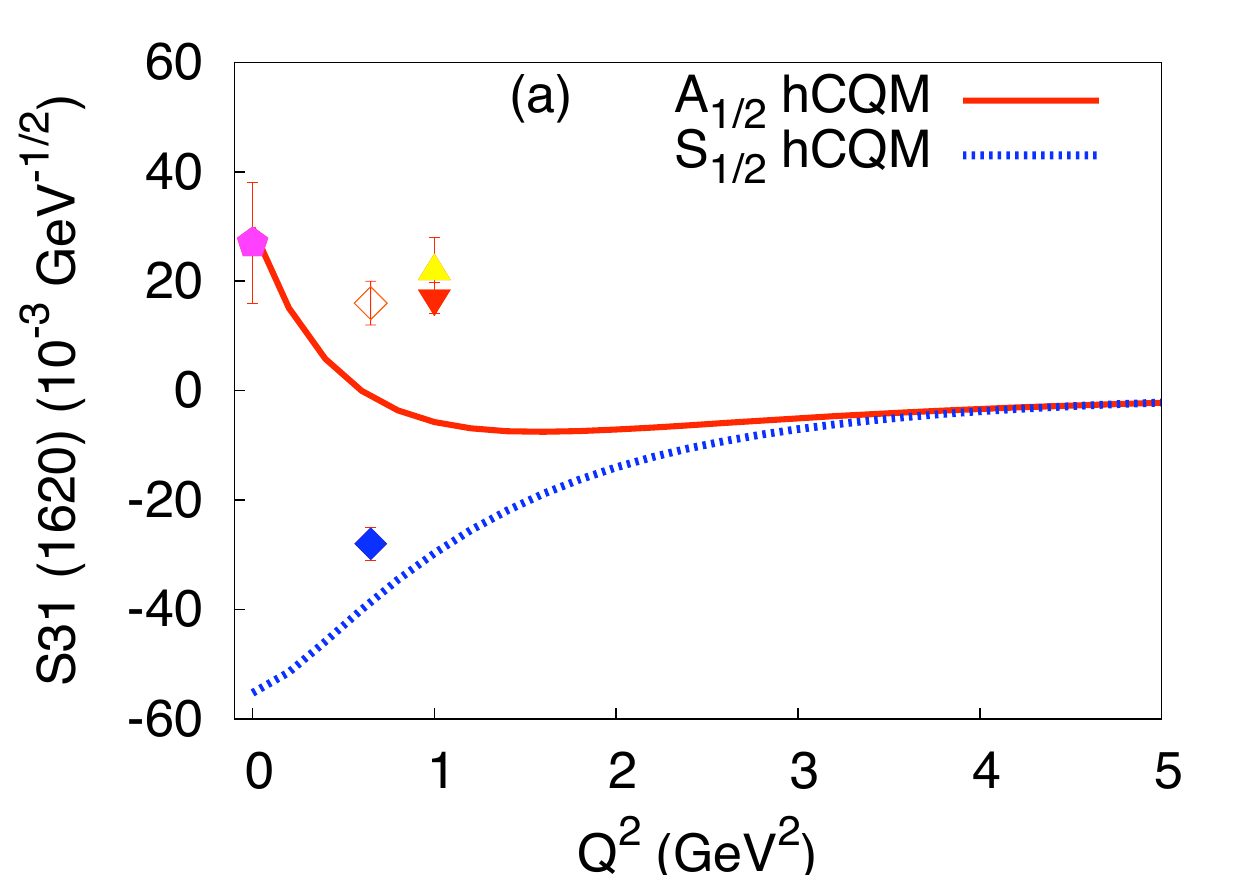}
\includegraphics[width=3in]{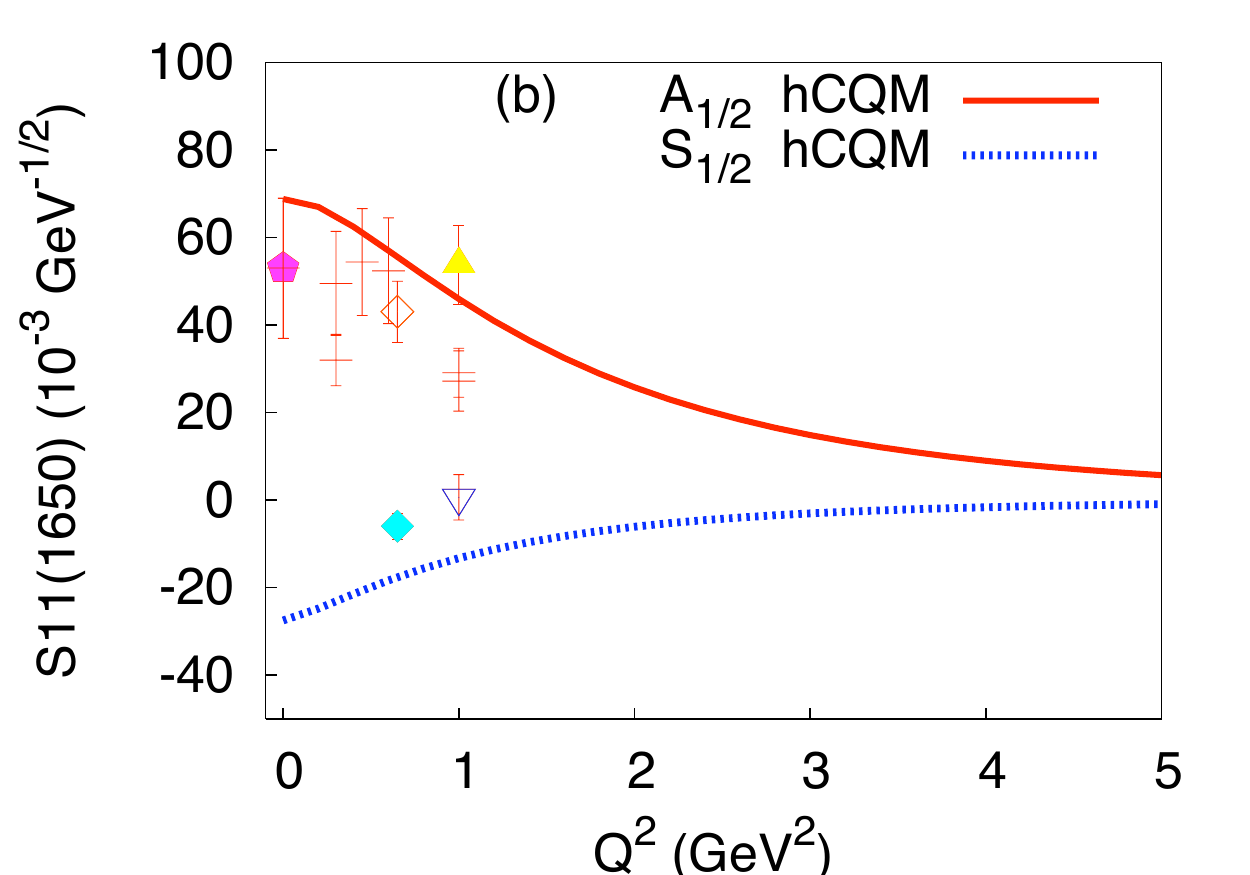}

\caption{(Color on line)The proton helicity amplitudes predicted by the hCQM  for the excitation of the S31(1620) (a) and S11(1650) (b), respectively, in comparison with the data of refs.  \cite{azn09} ($A_{1/2}$ open diamonds, $S_{1/2}$ full diamonds),\cite{azn05_2} ($A_{1/2}$ open diamonds, $S_{1/2}$ full diamonds), from the compilation reported in \cite{fh}  the Maid2007 analysis \cite{maid07} ($A_{1/2}$ up triangles, $S_{1/2}$ down triangles) of the data by refs. \cite{joo02} and \cite{lav04} and the compilation of the Bonn-Mainz-DESY data of refs. \cite{kru,bra,beck,breu} (crosses), presented in \cite{thom01}. The PDG points \cite{pdg} (pentagons) are also shown.}
\label{s}

\end{figure}
 
\begin{figure}[]
 
\includegraphics[width=3in]{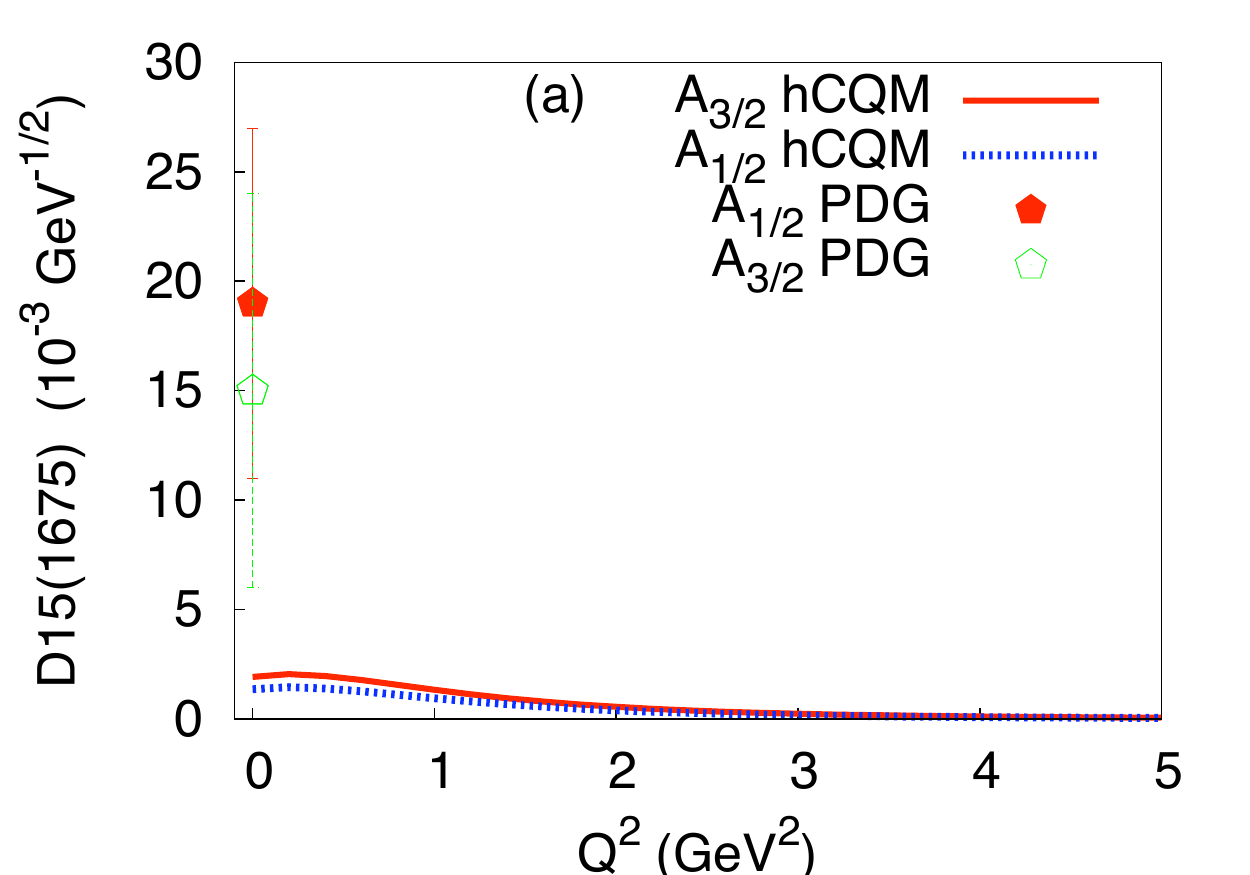}
\includegraphics[width=3in]{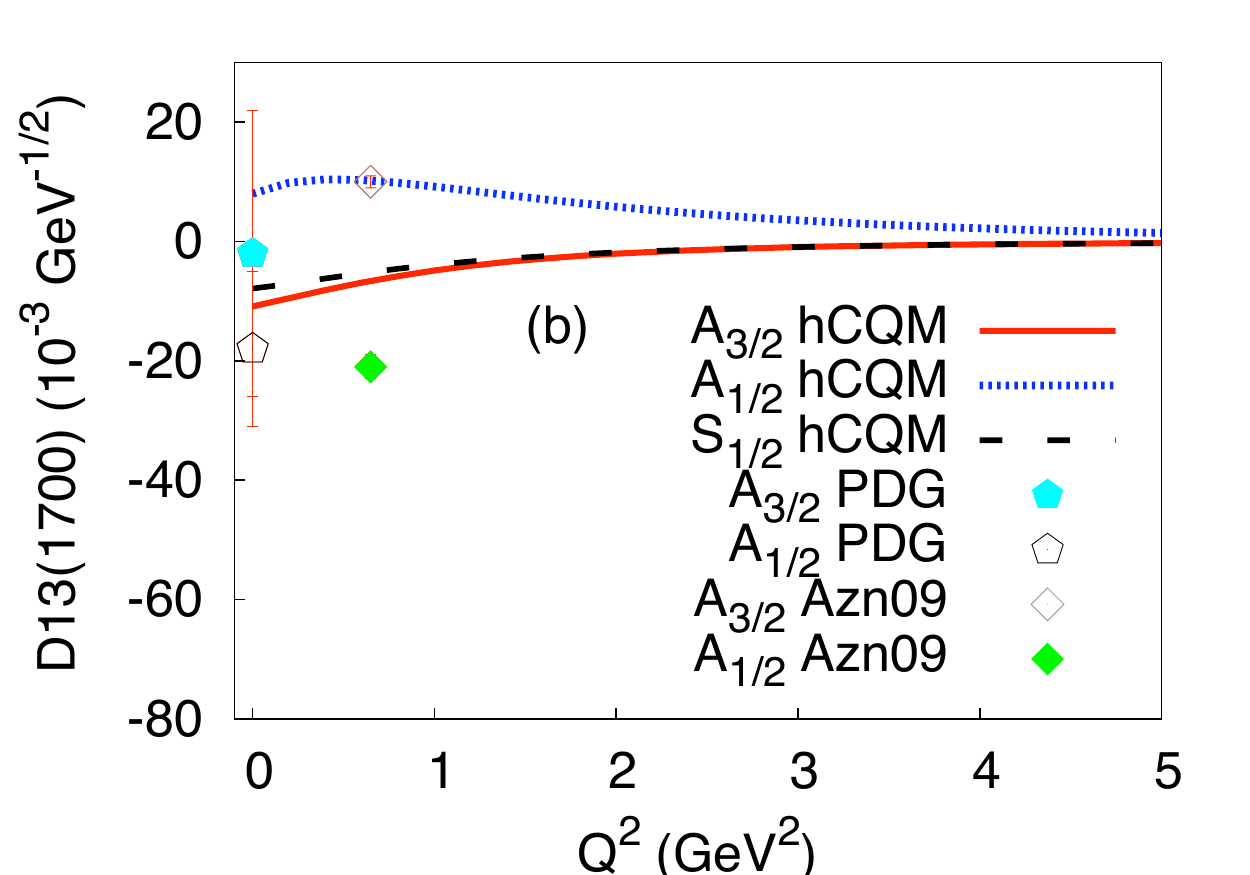}

\caption{(Color on line)The proton helicity amplitudes predicted by the hCQM  for the excitation of the  D15(1675) (a) and D13(1700) (b), respectively, in comparison with the data of refs.  \cite{azn09}. The PDG points \cite{pdg} are also shown.}
\label{d}
\end{figure}

The hCQM
results for the D13(1520) and the S11(1535) resonances 
\cite{aie2}, are given in Figs.(\ref{d13}) and (\ref{s11}), respectively. The agreement in the case of the
S11 is remarkable, the more so since the hCQM curve has been published three years in
advance with respect to the recent TJNAF data \cite{azn09}, \cite{azn05_1}, \cite{den07},  \cite{thom01}. 
In general the $Q^2$ behaviour is reproduced, except for
discrepancies at small $Q^2$, especially in the
$A^{p}_{3/2}$ amplitude of the transition to the $D_{13}(1520)$ state. 
These discrepancies could be ascribed either to the non-relativistic character of
the model or to the lack of explicit quark-antiquark configurations, which may be
important at low $Q^{2}$.  The kinematical relativistic corrections at the level of
boosting the nucleon  and the resonances states to a common frame are not 
responsible for these discrepancies,  as we have demonstrated in 
 Ref.\cite{mds2}.

\begin{figure}[h]

\includegraphics[width=3in]{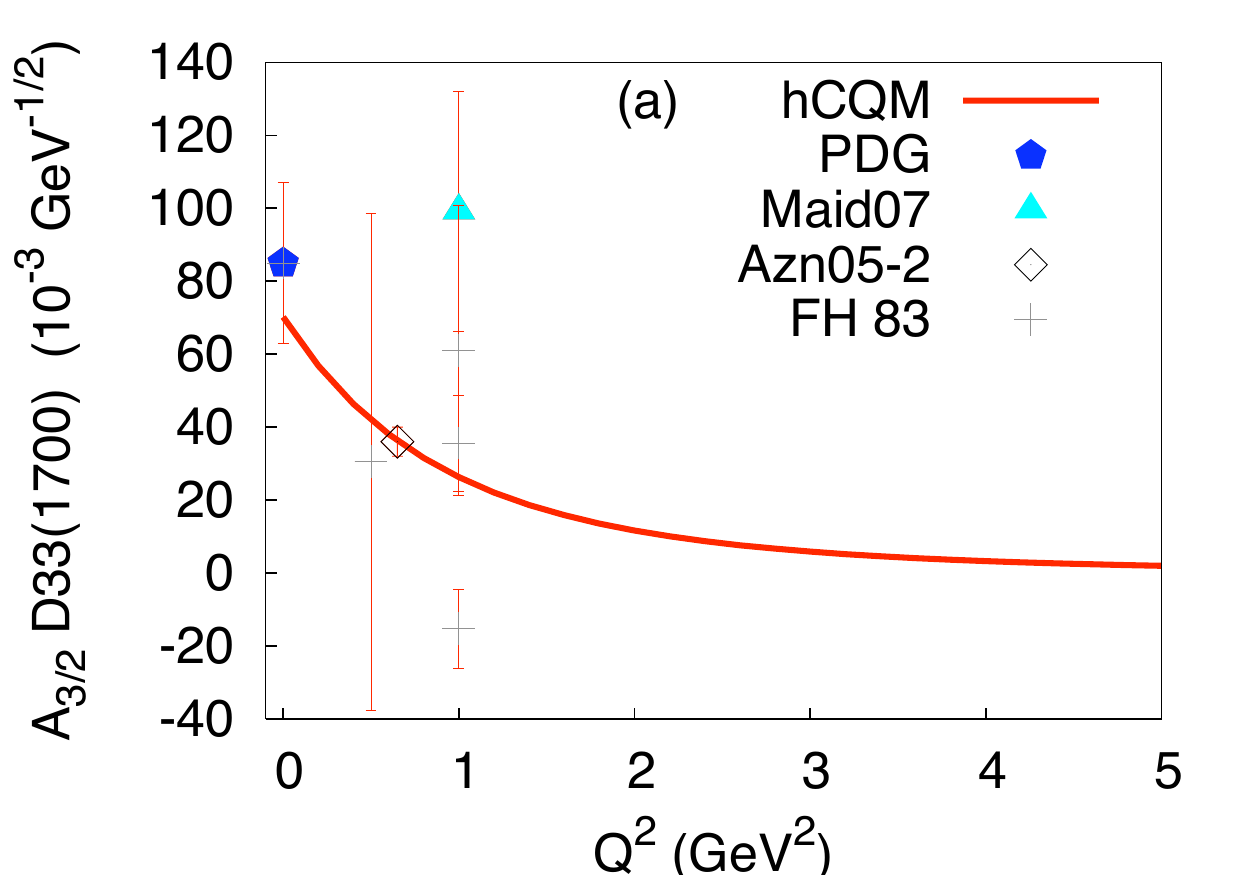} 
\includegraphics[width=3in]{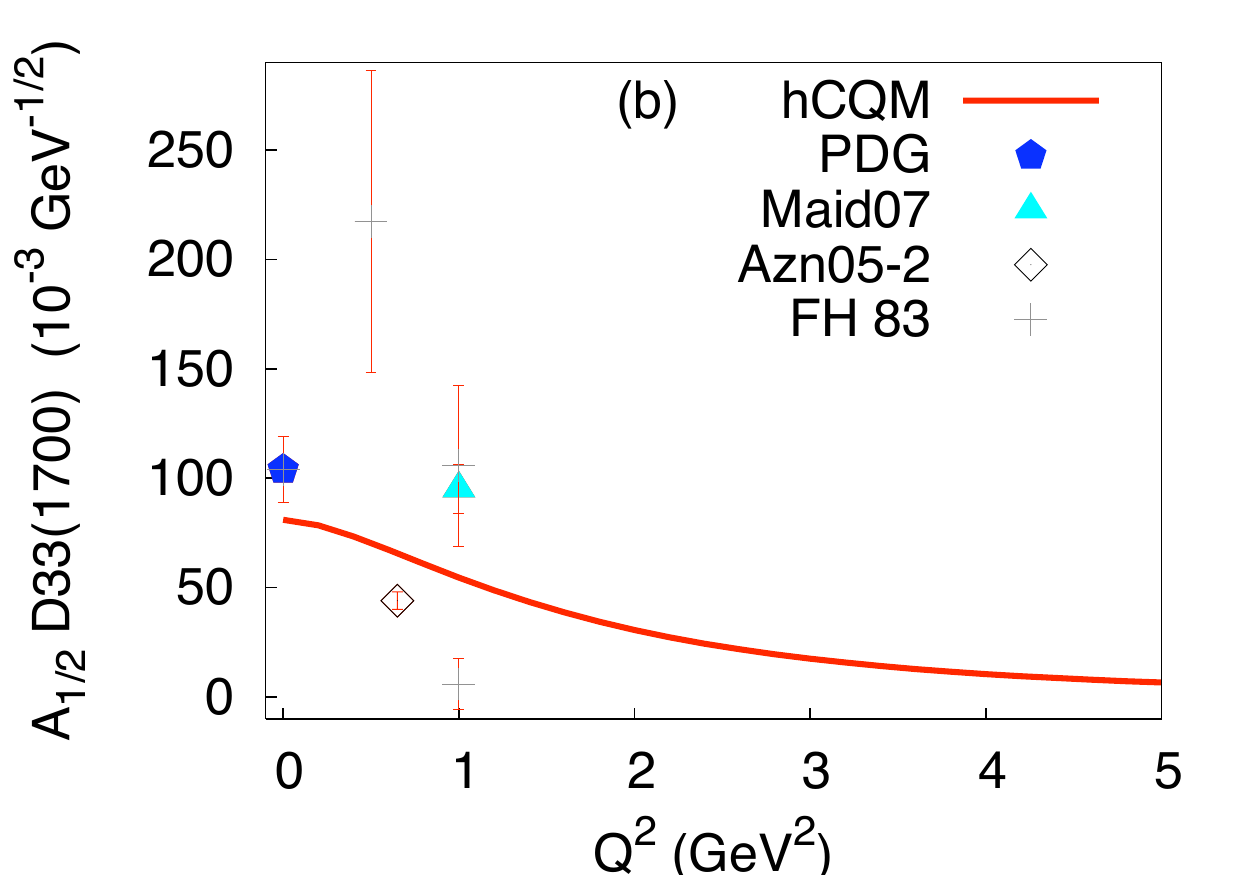}
\includegraphics[width=3in]{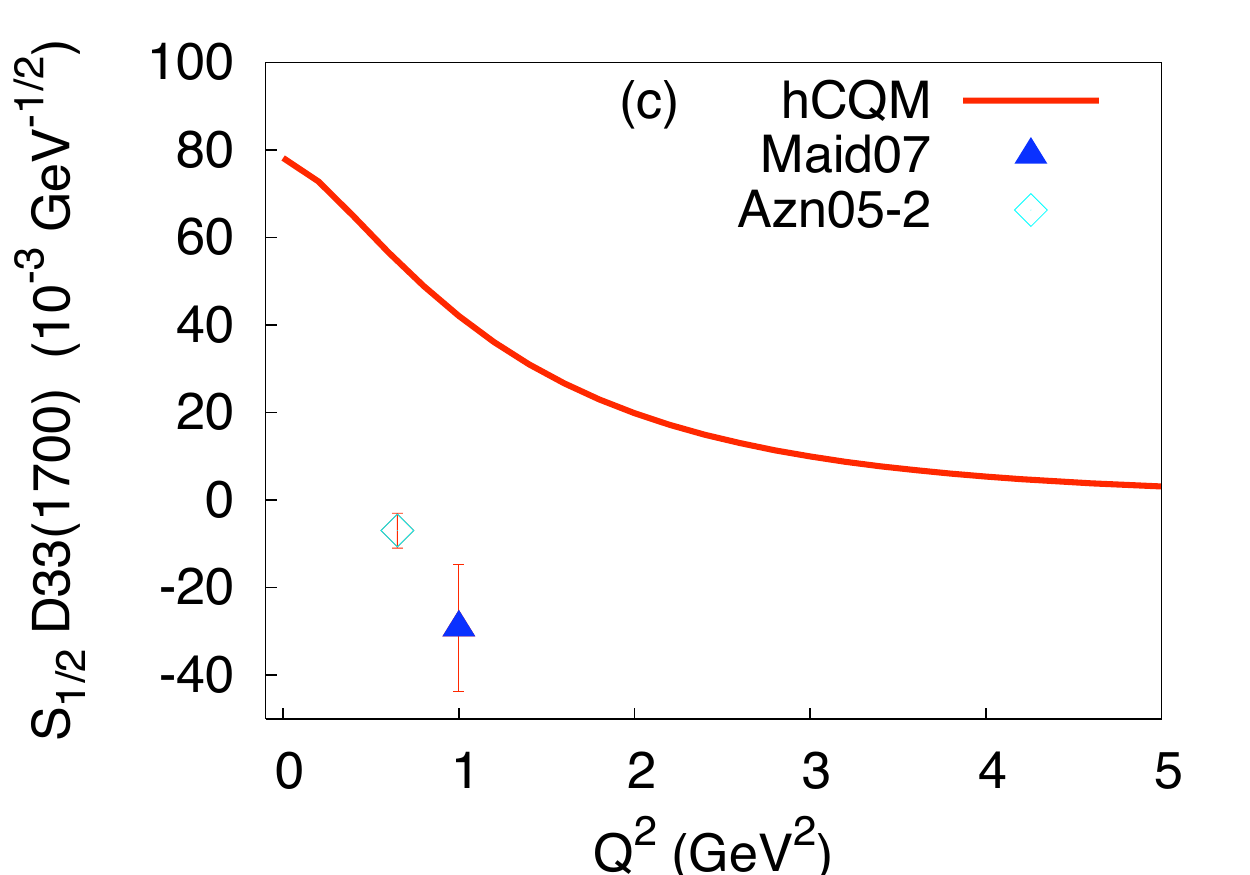}

\caption{(Color on line)The D33(1700) helicity amplitudes predicted by the hCQM (full curve)  $A_{3/2}$ (a), $A_{1/2}$ (b) and $S_{1/2}$ (c) in comparison with the data of \cite{azn05_2}  and the Maid2007 analysis \cite{maid07}  of the data by refs.  \cite{lav04}. The PDG points \cite{pdg} are also shown.}
\label{d33}
\end{figure}

 Similar results are obtained for the other negative
parity resonances  (see Figs.(\ref{s}, \ref{d}) and (\ref{d33})). 

It is interesting to discuss the influence of the hyperfine mixing on the excitation of the resonances. Usually there is quite a small difference between the values calculated with or without  hyperfine interaction. In some cases, however the excitation strength vanishes in the SU(6) limit and the non zero final result is entirely due to the hyperfine mixing of states. This happens for the excitation to the S11(1650) resonance, for both the transverse and longitudinal strengths.

The same thing happens for all the three helicity amplitudes of  the D13(1700) resonance, but in this case, at variance with the S11(1650) state,  the hyperfine mixing produces a low excitation strength. Also in the case of the transverse excitation of the D15(1675), the strength is given by the hyperfine mixing, while the longitudinal amplitude $S_{1/2}$ vanishes also in presence of a SU(6) violation.

It should be mentioned that the r.m.s. radius of the proton corresponding to the
parameters of Eq.(\ref{eq:par}) is $0.48~fm$, which is just the value fitted in
\cite{cko} to the $D13$ photocoupling. 
The missing strength at low $Q^2$ can be ascribed to the lack of
quark-antiquark effects \cite{aie2}, probably important in the outer region of the nucleon. In this way the emerging picture in connection with the resonance excitation is that of a small confinement zone of about $0.5 fm$ surrounded by a sort of  quark-antiquark (or meson) cloud.

For the higher negative parity resonance, the main problem is the lack of data, however the comparison with the hypercentral CQM do not contradict the observations made above.

\subsection{The proton excitation to the third and fourth resonance region}

\begin{figure}[h]

\includegraphics[width=3in]{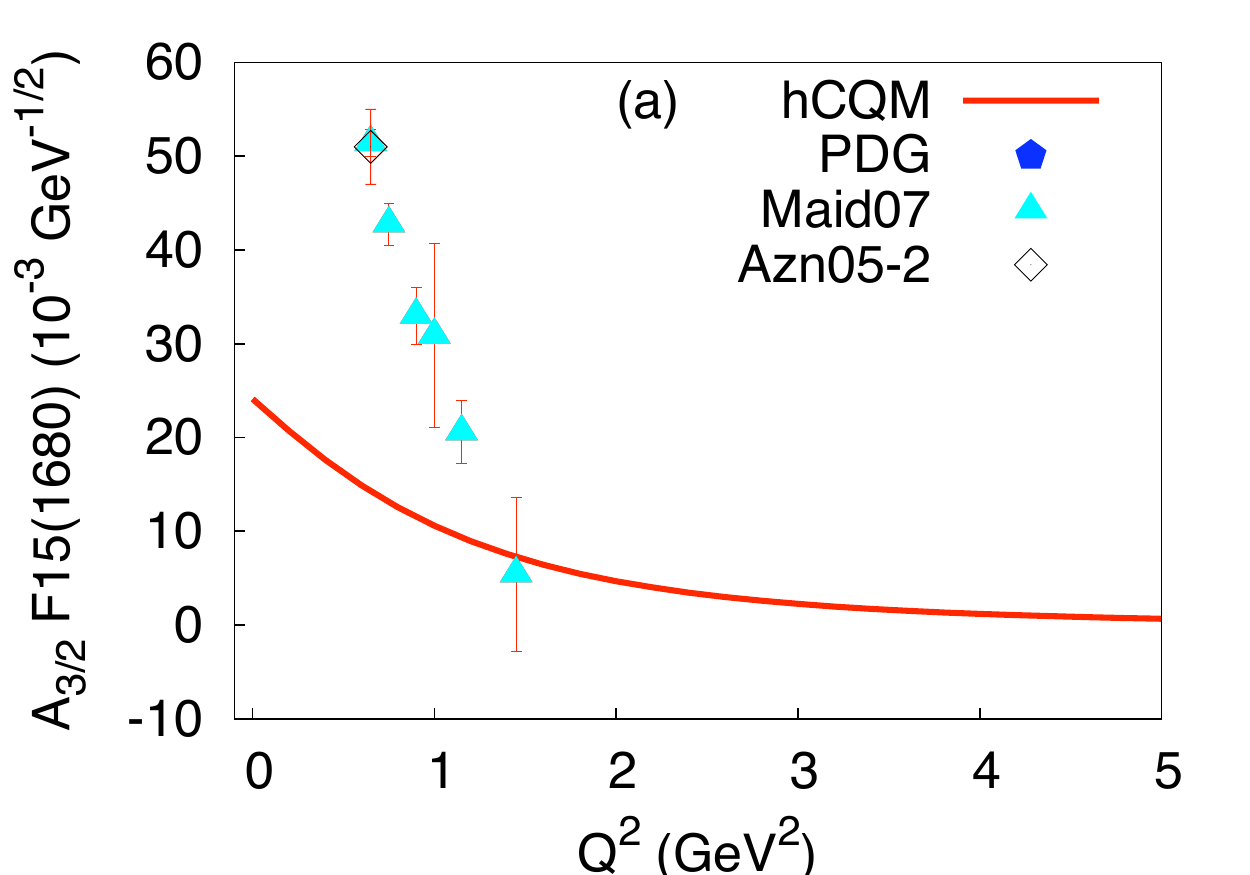} 
\includegraphics[width=3in]{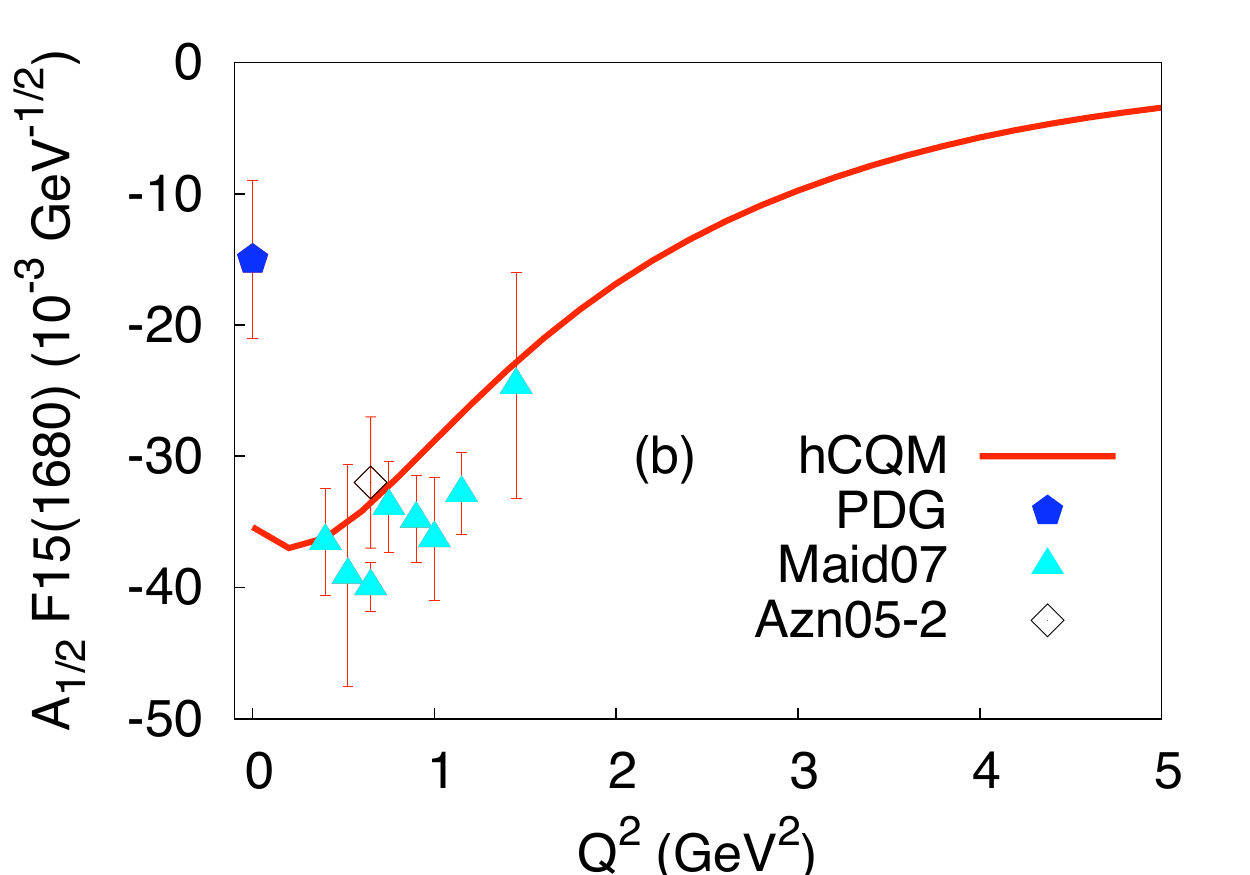}
\includegraphics[width=3in]{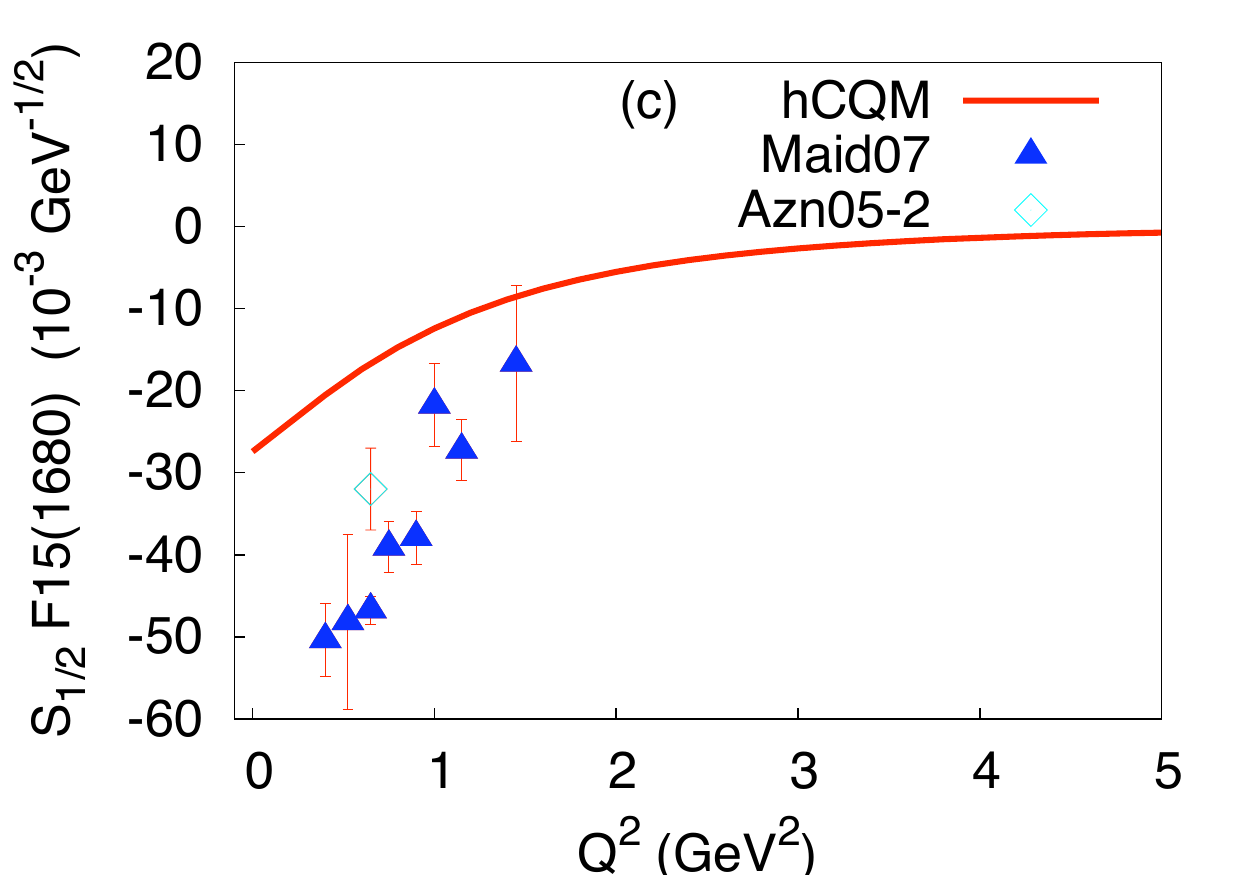}

\caption{(Color on line)The F15(1680) proton helicity amplitudes predicted by the hCQM  $A_{3/2}$ (a), $A_{1/2}$ (b) and $S_{1/2}$ (c) , in comparison with the data of refs.  \cite{azn05_2}  and the Maid2007 analysis \cite{maid07} of the data by refs.  \cite{lav04}. The PDG points \cite{pdg} are also shown.}
\label{f15}
\end{figure}

In this region the strength is dominated by the excitation to the F15(1680) resonance and the results calcualted with the hCQM are shown in Fig.(\ref{f15}).

\begin{figure}[h]

\includegraphics[width=3in]{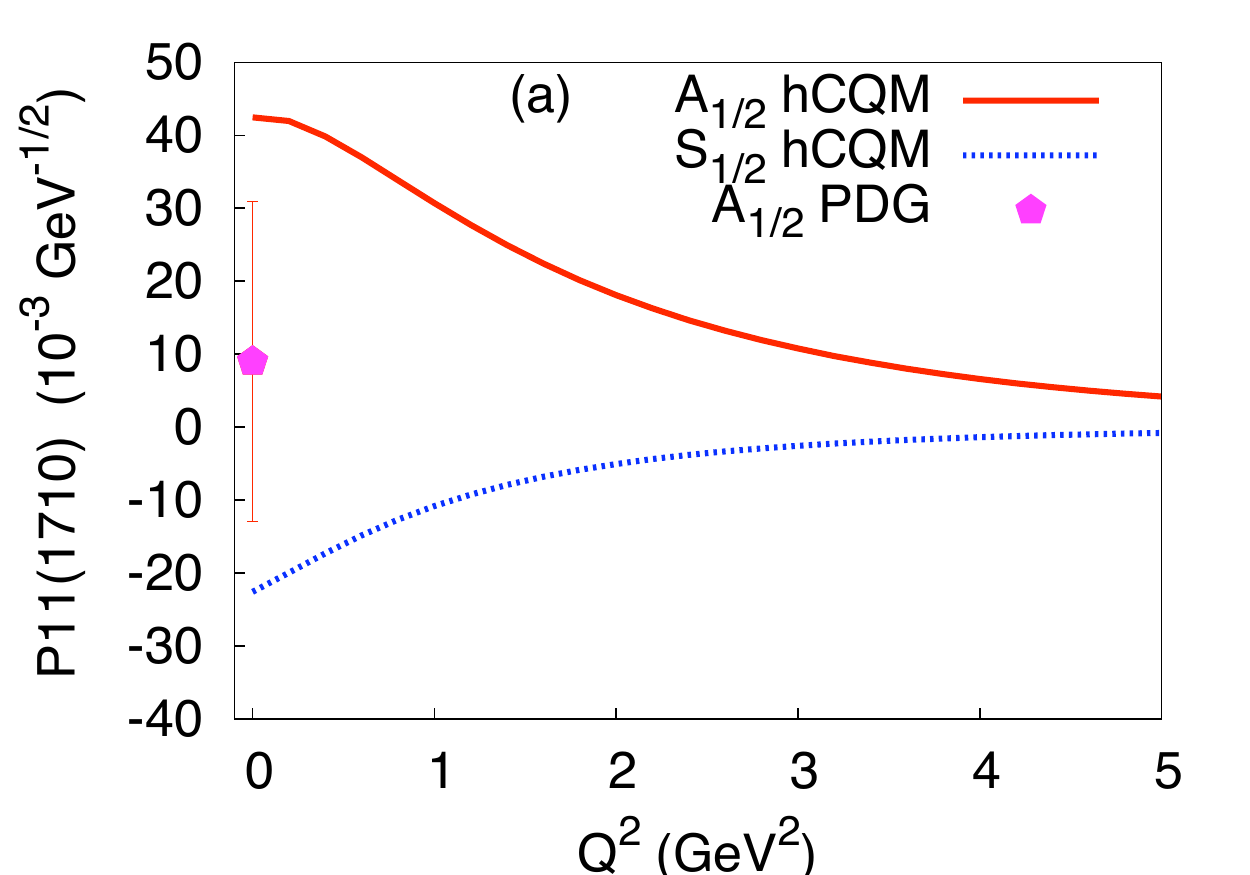}
\includegraphics[width=3in]{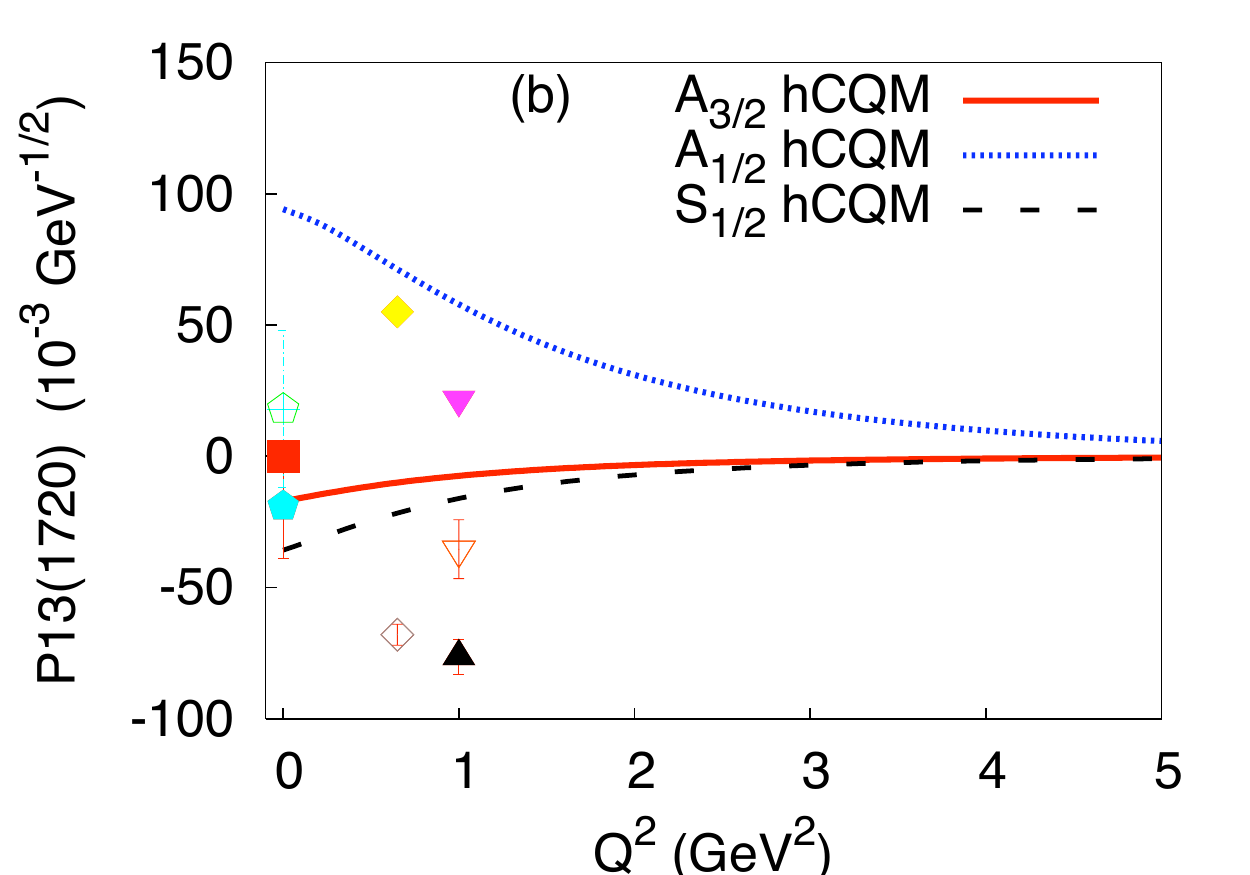}
\caption{(Color on line) The proton helicity amplitudes predicted by the hCQM  for the excitation of the P11(1710) (a) and P13(1720) (b),  respectively, in comparison with the data of ref. \cite{azn05_2} ($A_{3/2}$ open diamond, $A_{1/2}$ full diamond, $A_{3/2}$ full box) and the Maid2007 analysis \cite{maid07} ($A_{3/2}$ full up triangle, $A_{1/2}$ full down triangle, $A_{3/2}$ open down triangle) of the data by refs. \cite{joo02} and \cite{lav04}. The PDG point \cite{pdg} (pentagon) is also shown.
}
\label{p}

\end{figure}

\begin{figure}[h]

\includegraphics[width=3in]{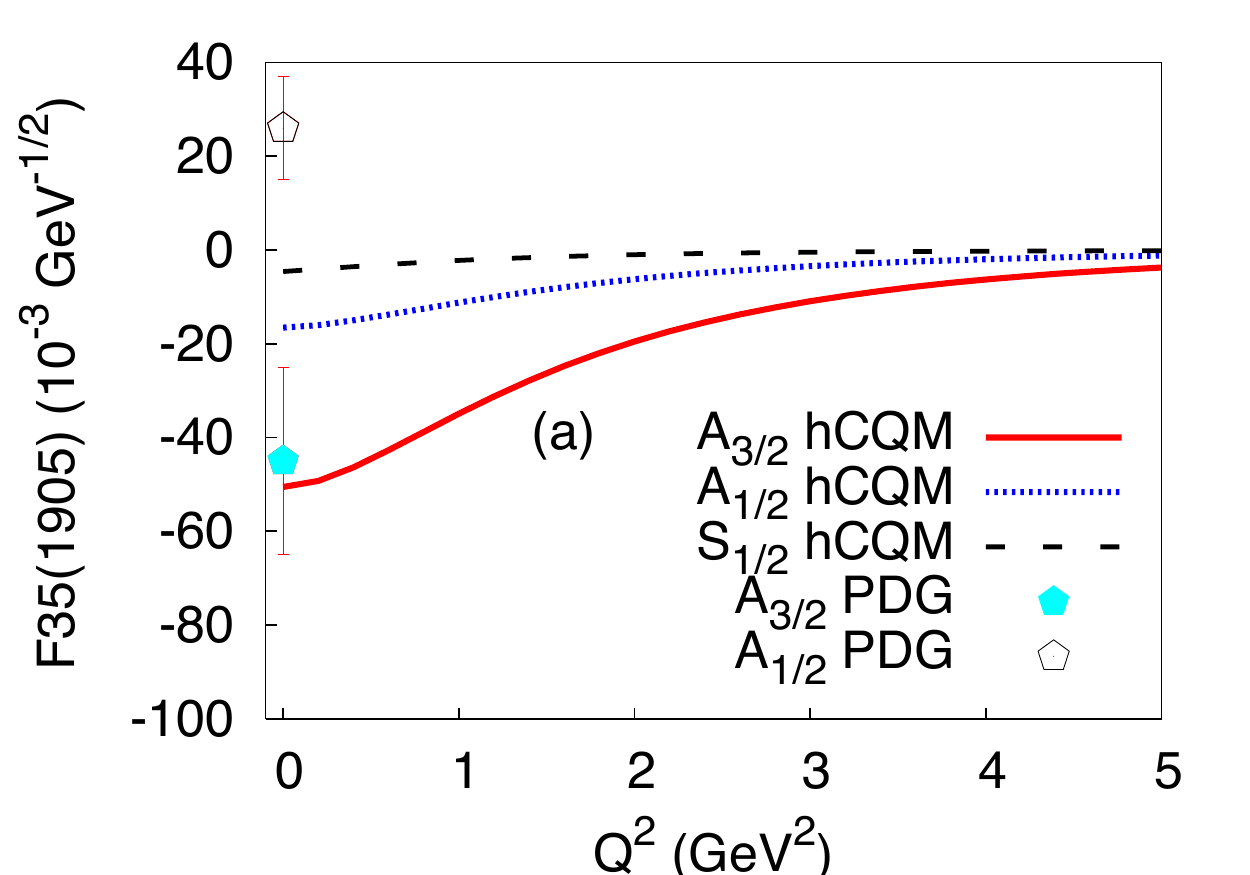} 
\includegraphics[width=3in]{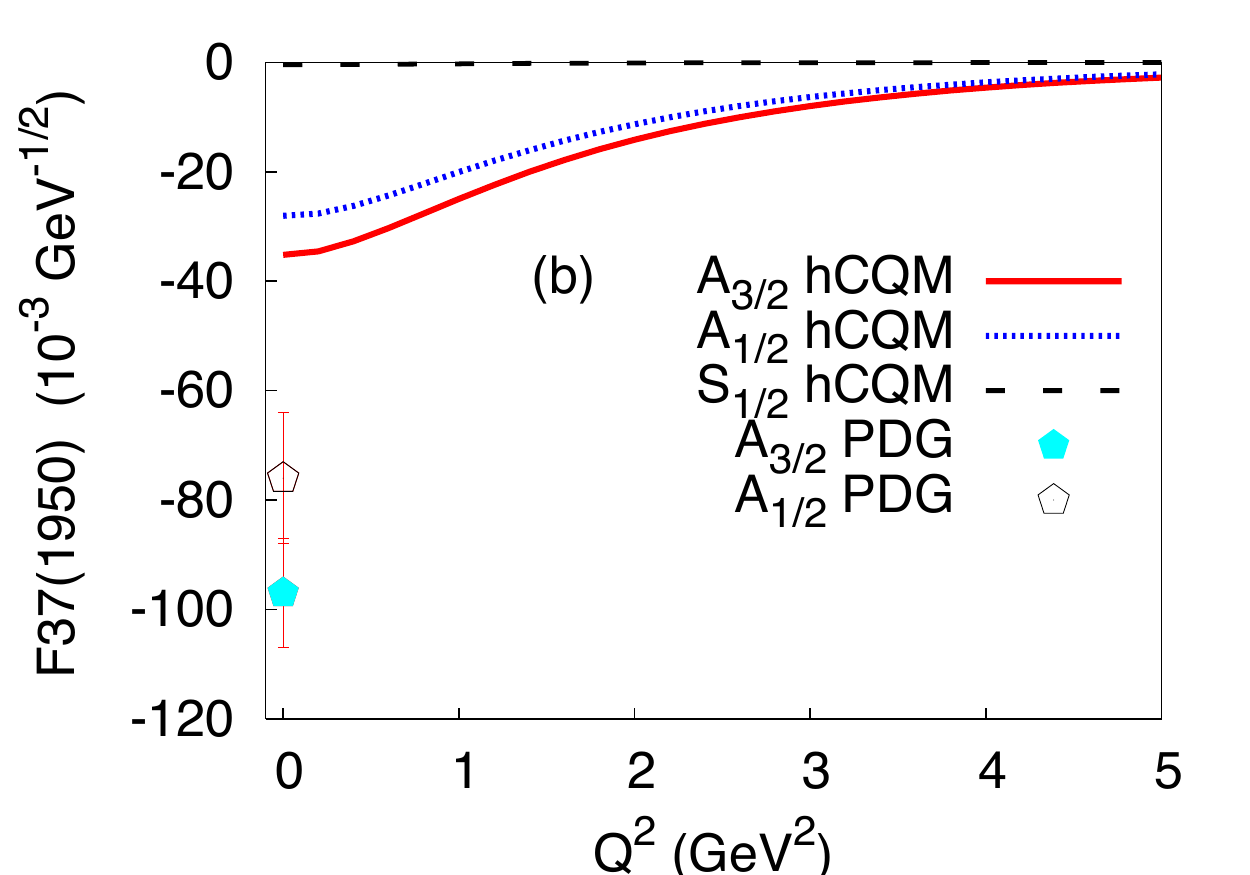}
\caption{(Color on line) The proton helicity amplitudes predicted by the hCQM  for the excitation of the F35(1905) (a) and F37(1950) (b), respectively. The PDG points \cite{pdg} (pentagons) are also shown.
}
\label{f}

\end{figure}

The situation is similar to that of the D13(1520) resonance. Here also the $A_{1/2}$ amplitude is fairly well reproduced, with possibly some problem in the low $Q^2$ region, while the $A_{3/2}$ amplitude exhibits a relevant lack of strength al low $Q^2$. For the longitudinal amplitude, the sign seems to be wrong, however new data at medium $Q^2$ are certainly needed.

As for the higher resonances, we note that our model predicts a second Roper-like state in the third resonance region (see Fig.(\ref{spect})). Performing the calculations identifying it with the state P11(1710), we get a relevant excitation strength (Fig.(\ref{pf})). The problem of a second Roper state is still open, since in some analysis there seems to be no evidence of its existence \cite{pdg,arn06}; on the contrary, the presence of a state P11(1710) is supported by the recent analysis based on the Dubna-Mainz-Taipei (DMT)
dynamical model \cite{dmt2}. In this energy region, the strength seems to be dominated by the P13(1720) resonance (Fig.(\ref{pf})), for which few data up to $Q^2 = 1 (GeV)^2$ \cite{azn05_2} are available. With these scarse data a comparison is preliminary, however here again there seems to be a lack of strength specially for the $A_{3/2}$ amplitude.

In Fig.(\ref{pf}), the results for the remaining higher resonances F35(1905) and F37(1950) are shown. For the former, a relevant excitation strength is predicted for the $A_{3/2}$ amplitude. Hopefully, new data will allow a reasonable comparison with the theoretical quantities.

\subsection{The neutron excitation to the $I=\frac{1}{2}$ states }

Of course, the proton and neutron excitations to the $I=\frac{3}{2}$  states are the same for isospin reasons. 

\begin{figure}[h]
\includegraphics[width=3in]{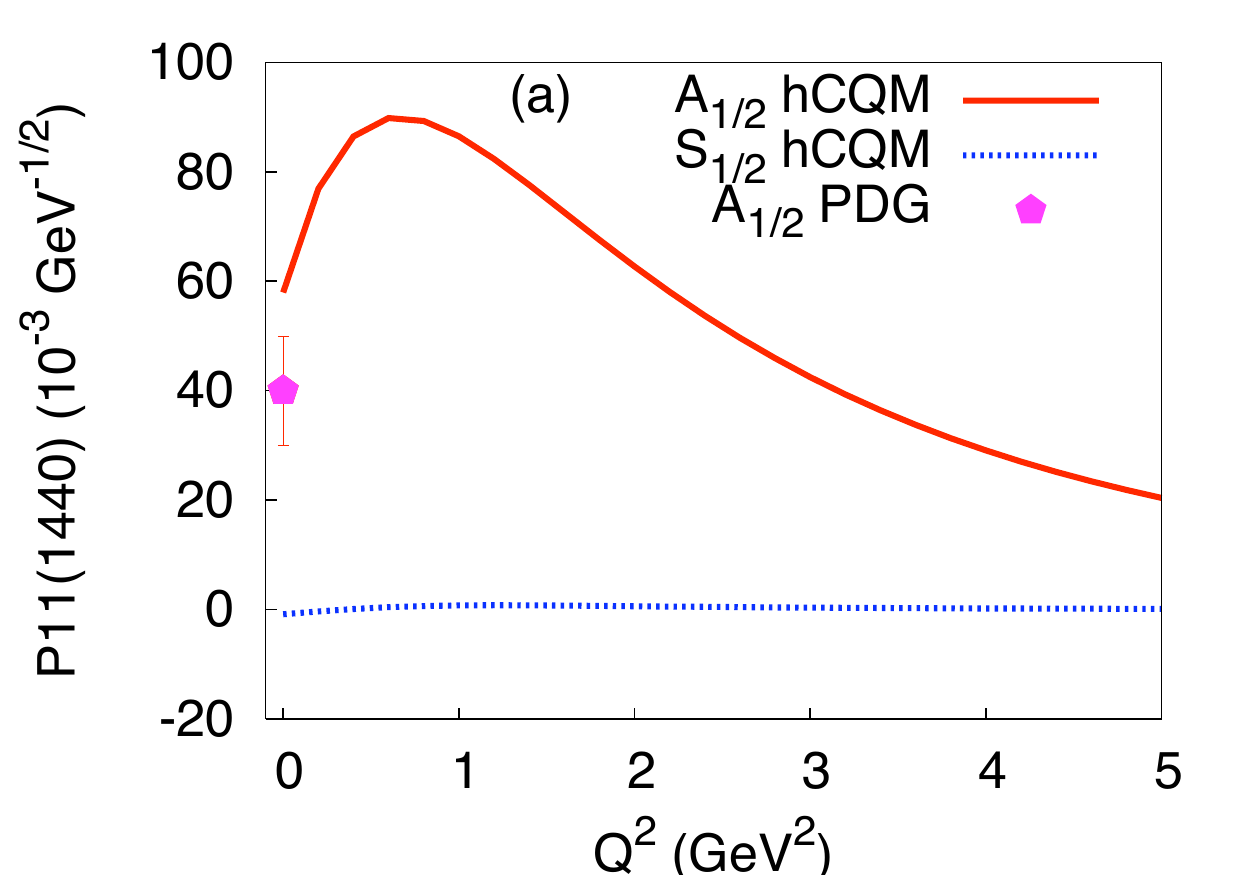}
\includegraphics[width=3in]{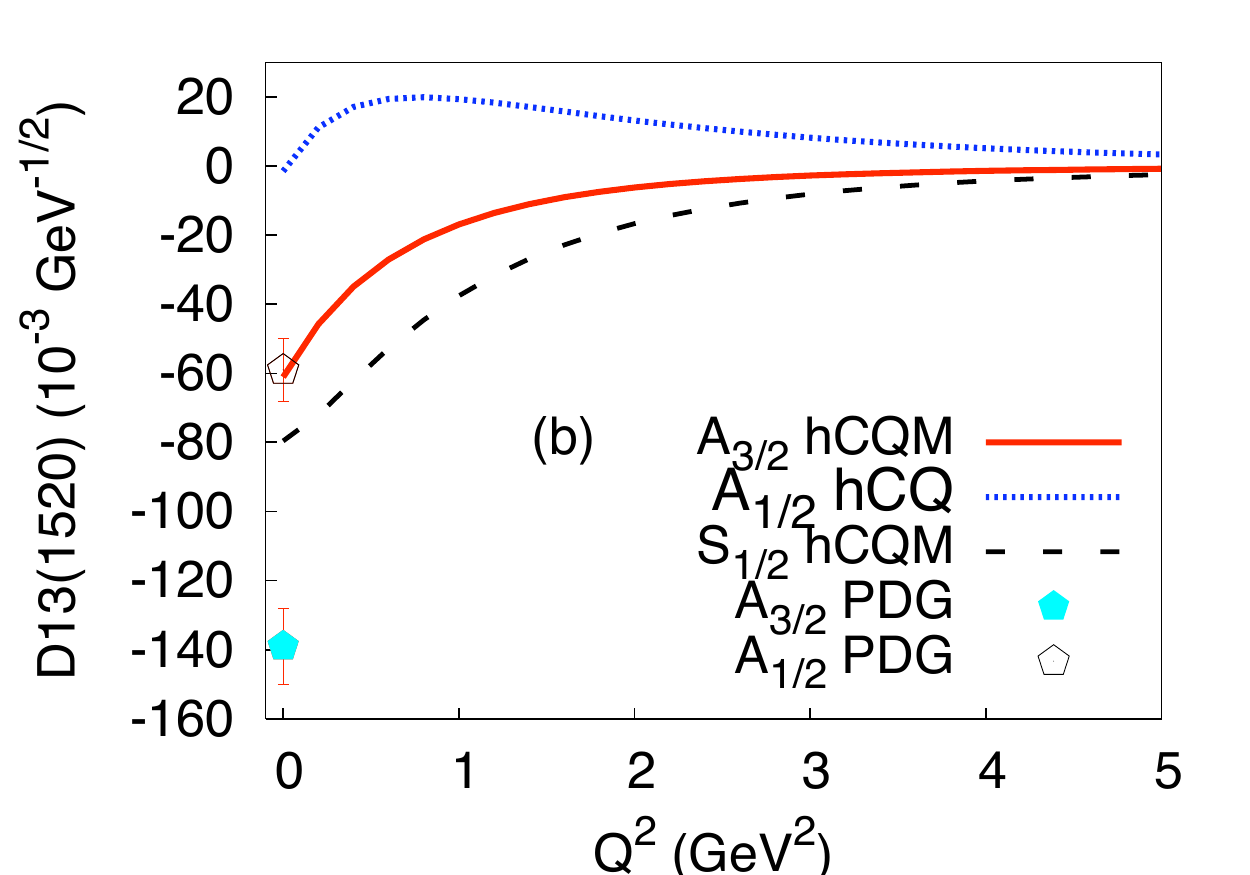}
\includegraphics[width=3in]{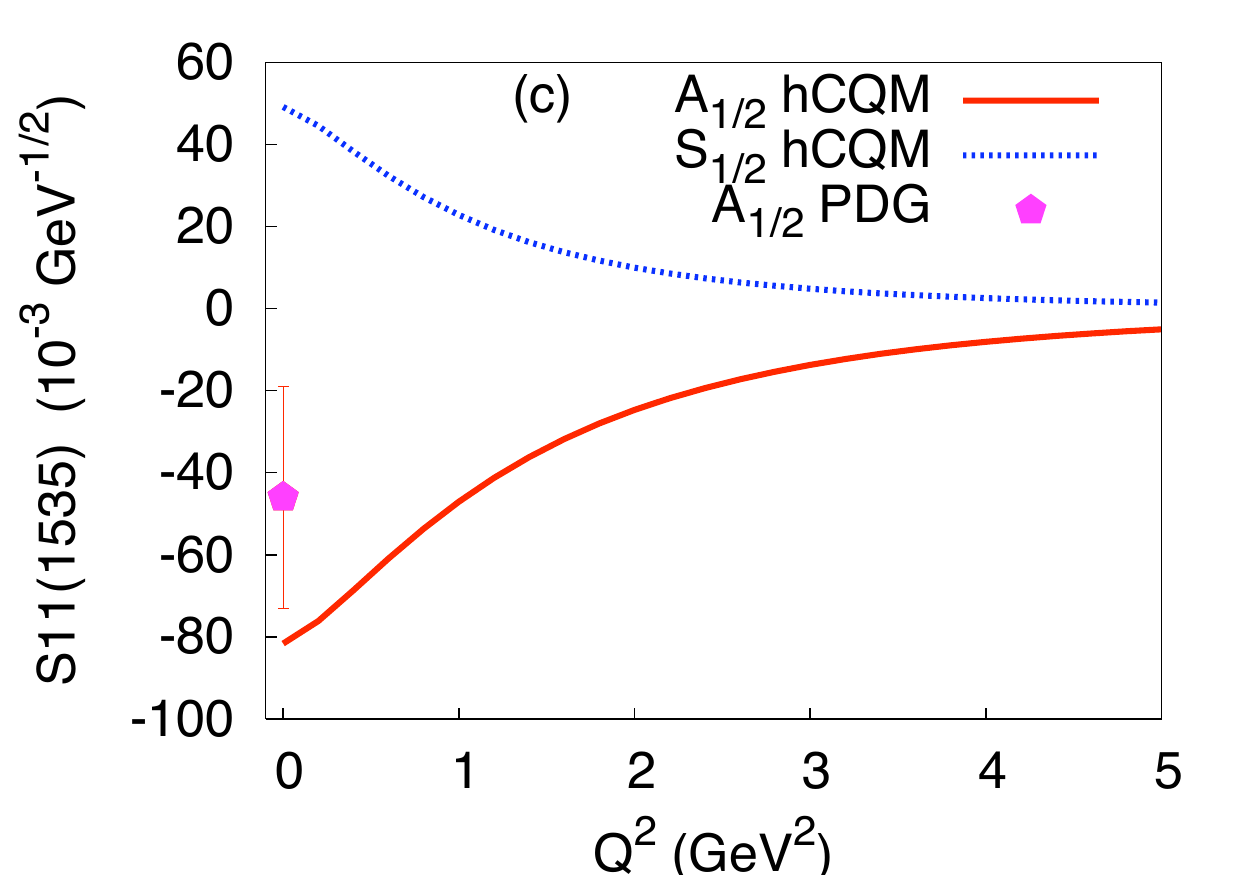}

\caption{(Color on line) The neutron excitation strength predicted by the hypercentral CQM, in comparison with the PDG points  \cite{pdg} (Part I).}
\label{n1}

\end{figure}

\begin{figure}[h]
\includegraphics[width=3in]{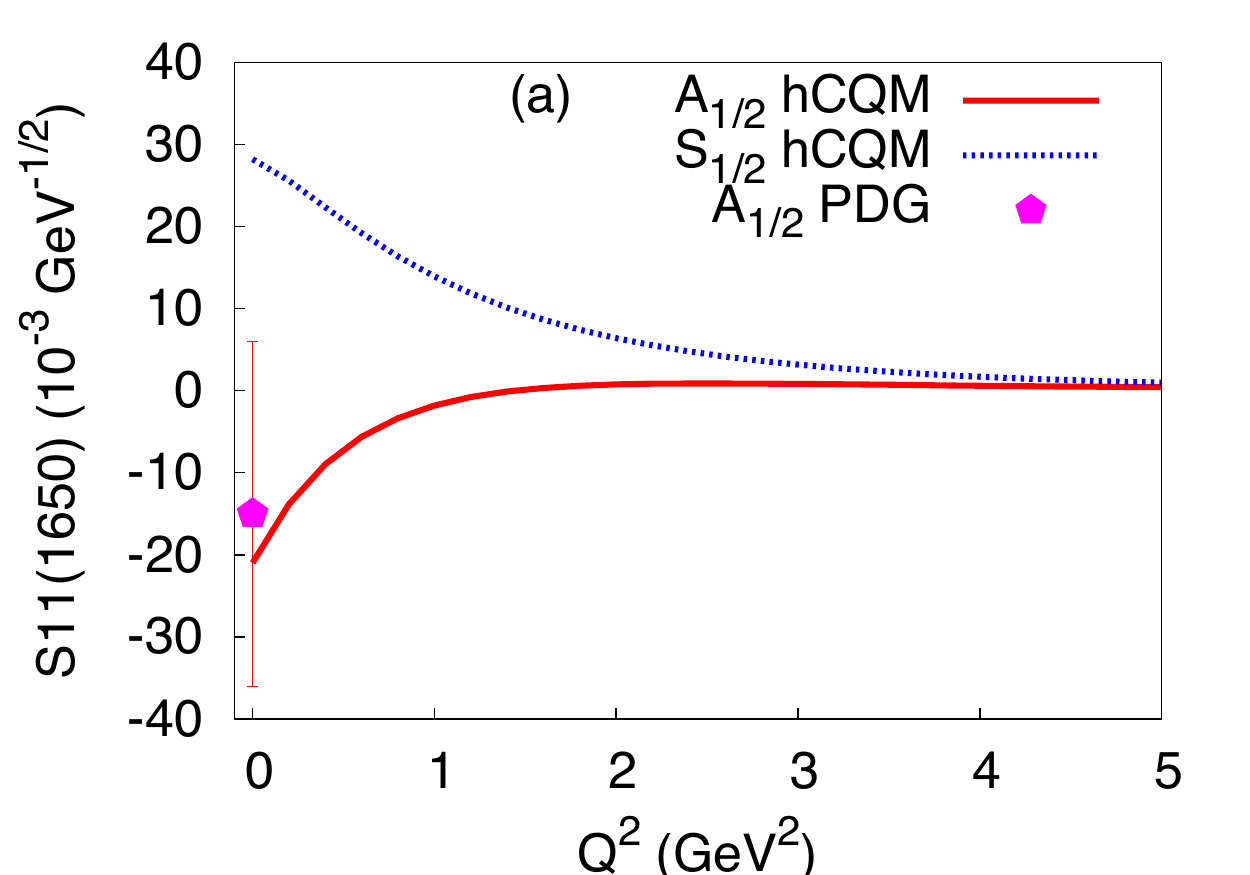}
\includegraphics[width=3in]{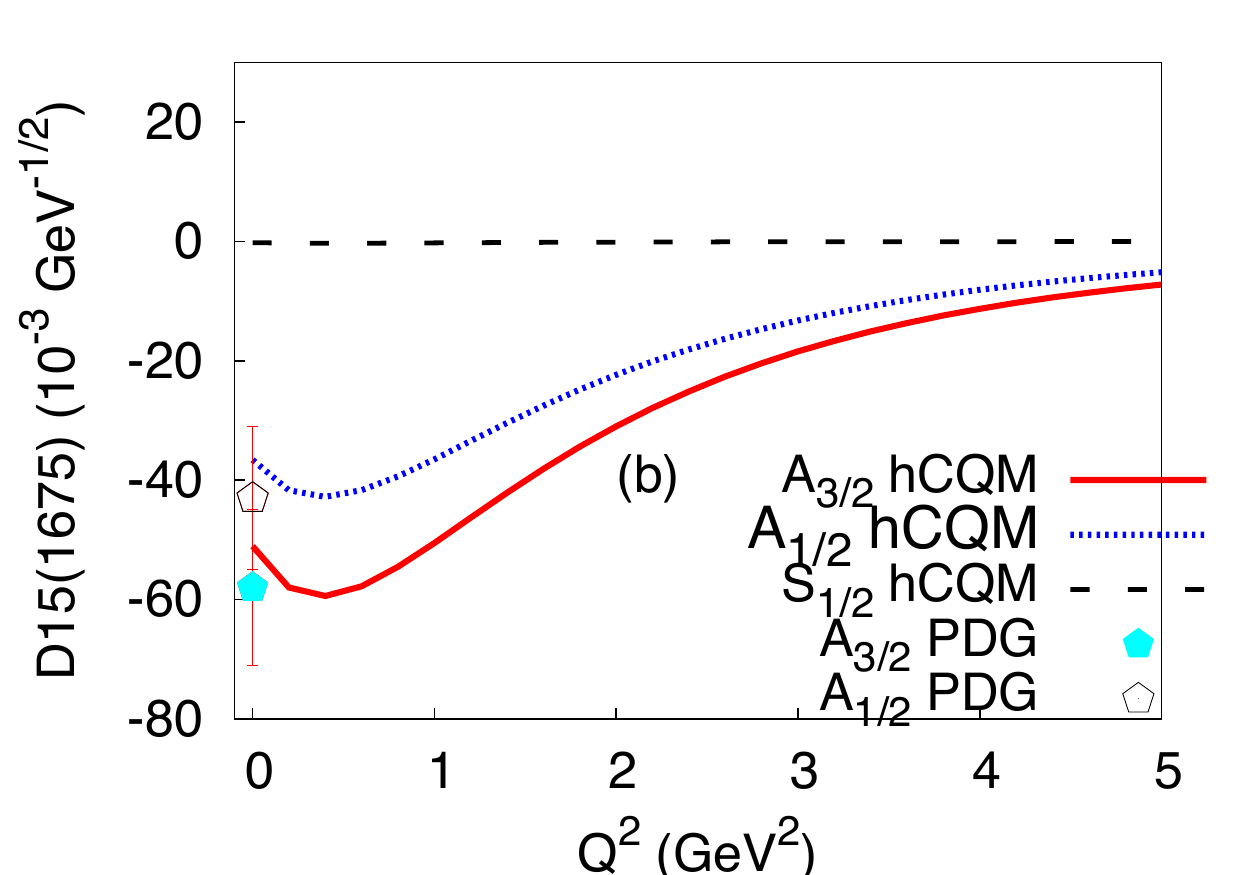}
\includegraphics[width=3in]{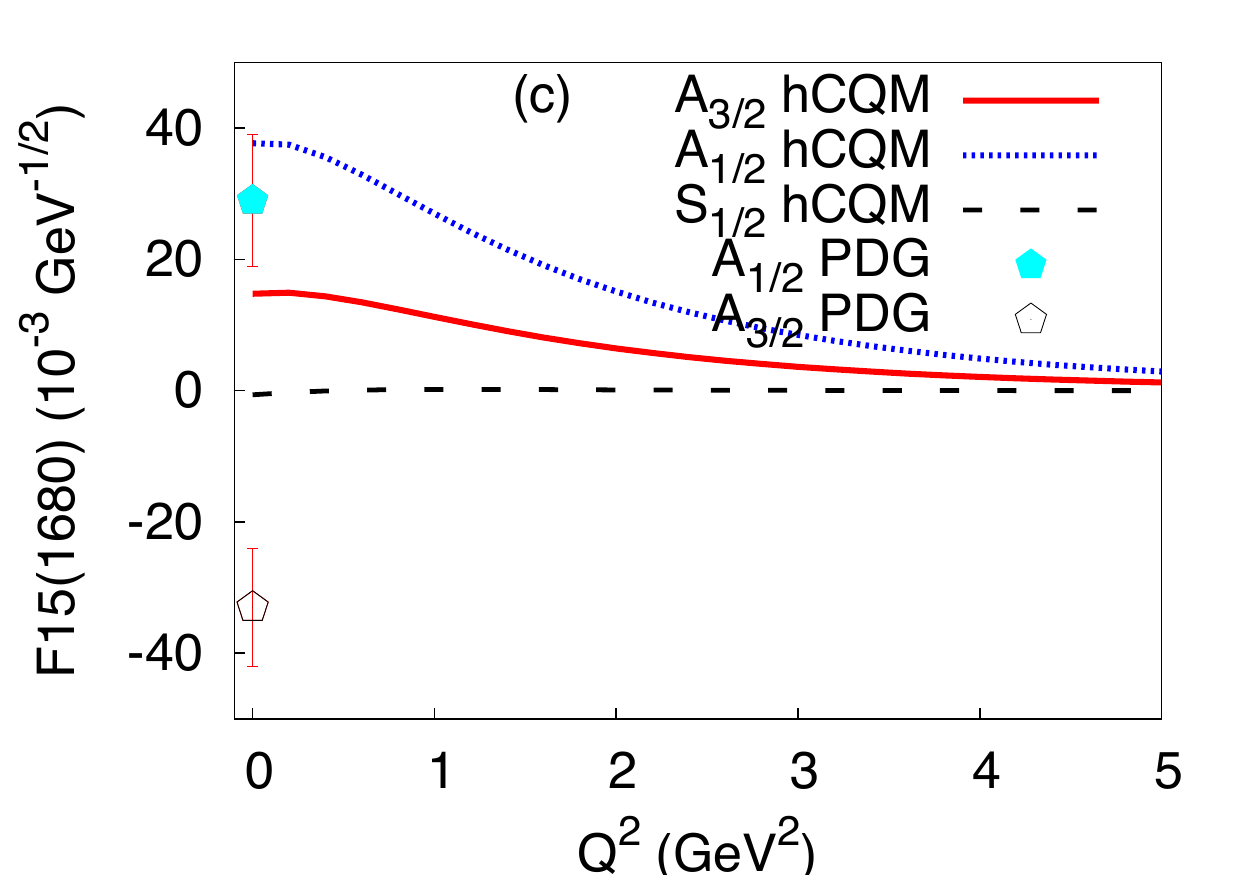}

\caption{(Color on line) The neutron excitation strength predicted by the hypercentral CQM, in comparison with the PDG points  \cite{pdg} (Part II).}
\label{n2}
\end{figure}

\begin{figure}[h]
\includegraphics[width=3in]{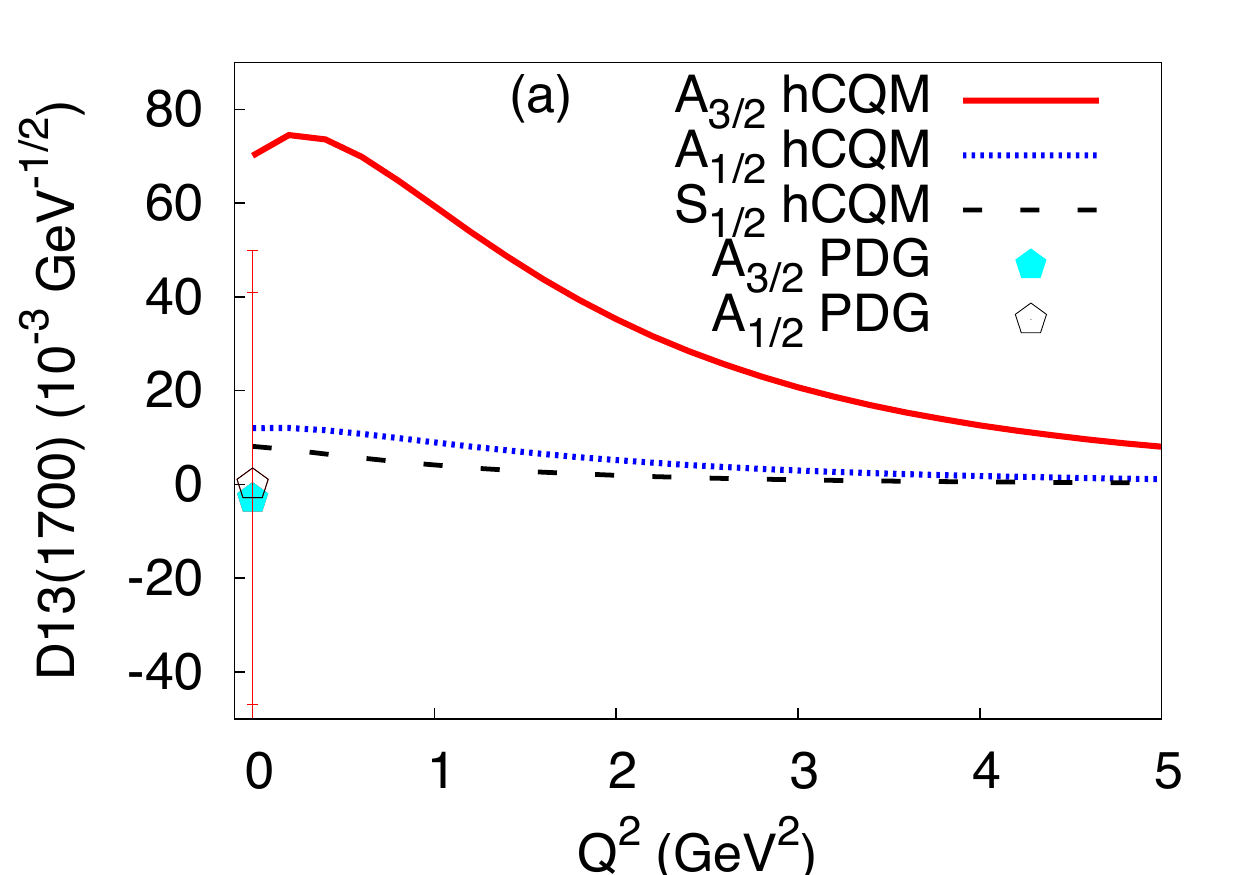}
\includegraphics[width=3in]{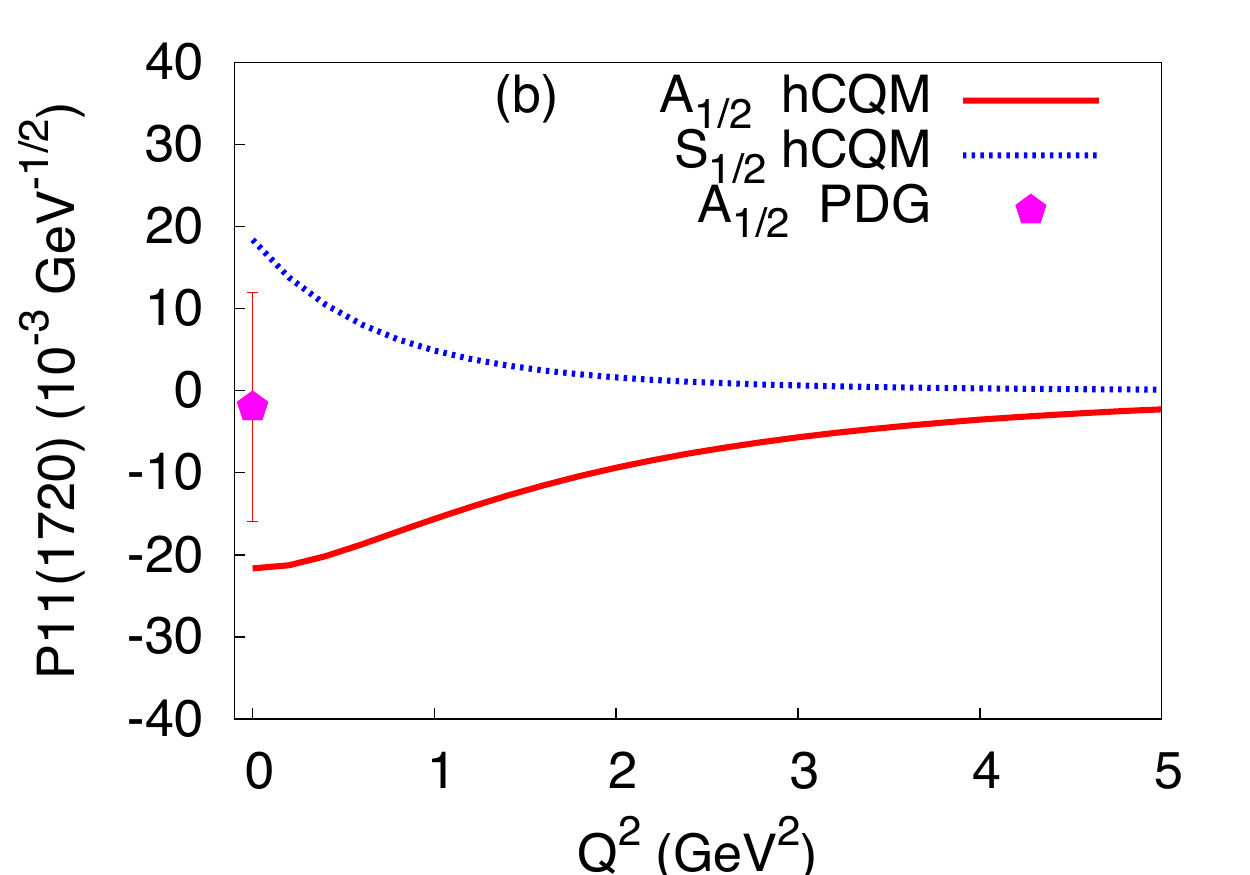}
\includegraphics[width=3in]{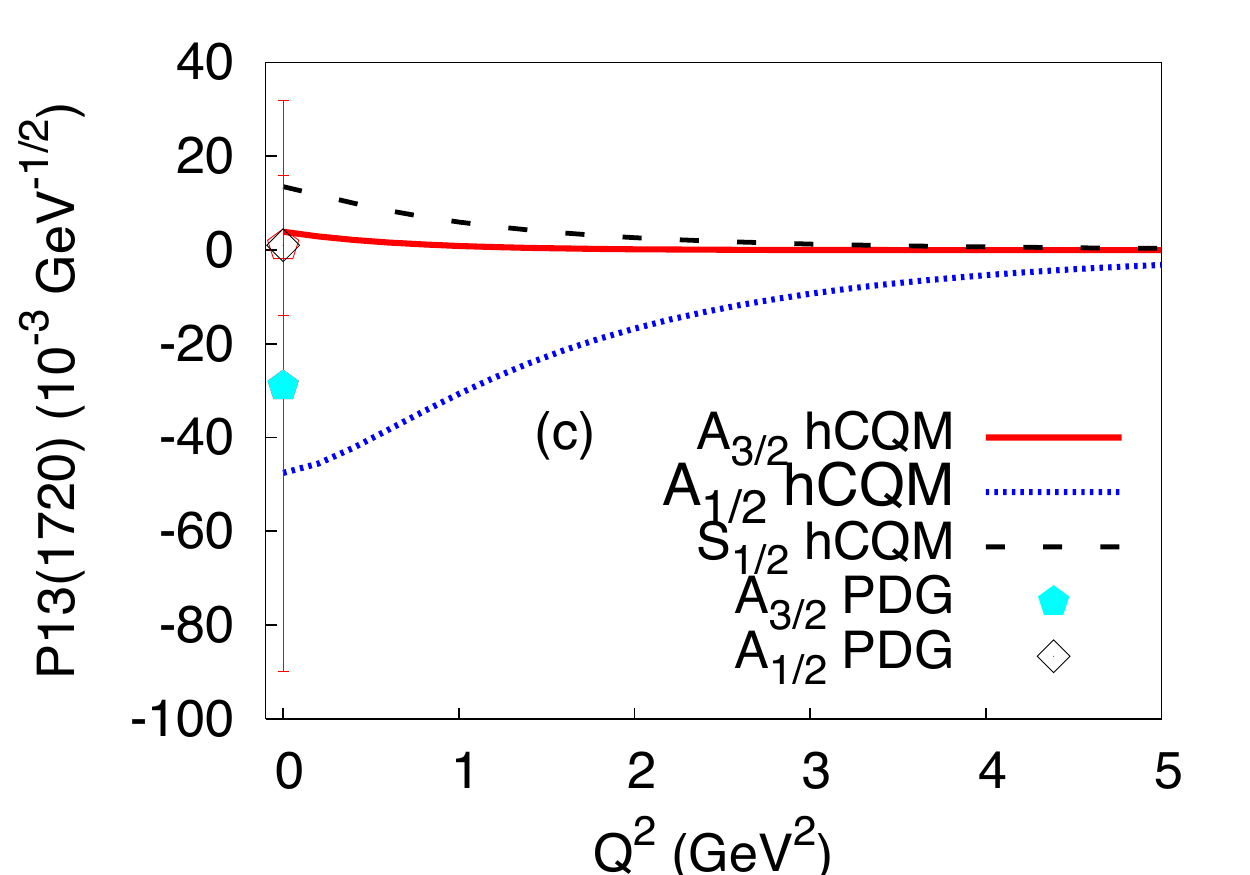}

\caption{(Color on line)The neutron excitation strength predicted by the hypercentral CQM, in comparison with the PDG points  \cite{pdg} (Part III).}
\label{n3}

\end{figure}

In all the other cases there are both isoscalar and isovector contributions, which make the neutron strength quite different from the proton ones, as it can be seen in Figs.(\ref{n1}, \ref{n2} and \ref{n3}). 

As a general observation, at least one of the neutron helicity amplitudes is not negligible. Also the longitudinal (charge) excitation is often relevant, an effect due clearly to the presence of charged particles moving within the neutron. Particularly interesting is the comparison between the two Roper resonances: for the P11(1440) state, the excitation is purely transverse, while the P11(1710) presents comparable longitudinal and transverse amplitudes, although opposite in sign.

The measurement of the neutron excitation is difficult, since one has to rely upon targets with bound neutrons, but an experimental information would be highly important in order to test our knowledge on the internal nucleon structure.

\section{Conclusions}

We have reported the predictions of the hCQM \cite{pl} for the transverse and longitudinal helicity amplitudes regarding the electromagnetic excitation of various baryon resonances. The transverse excitation amplitudes to the negative parity resonances have been already presented in \cite{aie2}. The excitations chosen in this paper concern the 12 baryon resonance which, according to the PDG classification \cite{pdg},  exhibit a relevant photoexcitation and, in addition, the two resonances D13(1700) and P13(1720), which seem to play a relevant role in the phenomenological analyses.

The comparison between data and theoretical results suffers from the lack of experimental points for various resonances, specially at high $Q^2$. However, the model, even if  non relativistic, is able to give an overall description of present data, in particular of the medium-high $Q^2$ behavior of many amplitudes. At low $Q^2$ there is often a lack of strength, specially for the transverse $A_{3/2}$ amplitudes. We recall that the calculated proton radius provided by the model is about $0.48 fm$ and that a radius of this size is required for the description of the
e.m. excitation to the $F15$ and $D13$ resonances \cite{cko}.
The smallness of the proton radius  together with the lack of strength in the low $Q^2$
region, suggest an interesting picture for the proton (and consequently
for hadrons), namely that of a small core, with radius of about
$0.5÷fm$, surrounded by an external quark-antiquark (or meson) cloud. 
The contributions coming from this external cloud have been pointed out as a possible origin of the 
missing strength \cite{es,aie,aie2} and are  obviously lacking in the
available CQMs. Their effect is expected to decrease for medium-high $Q^2$ and
therefore it is not a surprise that CQMs fail to reproduce the strength at
the photon point but give reasonable results for medium $Q^2$. These
considerations are supported by the evaluation of the pion contributions performed using the Mainz-Dubna dynamical model \cite{dmt}, which have shown that the importance of such contributions systematically decreases with increasing $Q^2$, going rapidly to zero \cite{ts03}. 

Another important issue is relativity. The present hCQM is non relativistic and allows to calculate the three quark wave function in the baryon rest frame. The e.m. form factors are evaluated in the Breit frame (see Eq.(\ref{eq:breit})) and this implies that a Lorentz transformation should to be applied to both the initial and final baryon state. This can be done quite easily \cite{mds2}, with the result that the helicity amplitudes are only slightly modified, probably because the high mass values of the resonances lead to a non relevant recoil of the three quark states. Such situation is quite different from the elastic form factor case \cite{mds}, where the simple application of Lorentz boosts is important but still not sufficient to obtain a good description of the experimental data: in fact a fully relativistic theory is needed, as in refs. \cite{card_ff,wagen,boffi,mert,ff_07,ff_10}, which use a relativistic hamiltonian for the dynamics of the three quark system. Such relativistic formulation should be extended also to the calculation of the helicity form factors (work in this direction is in progress), however, because of the above considerations,  the expectation is that the calculated helicity amplitudes should not differ too much from the non relativistic ones and that the relativistic corrections are not responsible for the lack of strength at low $Q^2$.

Finally it should be reminded that a good description of the elastic form factors in some cases is achieved only introducing intrinsic quark form factors \cite{card_ff,ff_07,ff_10}. Presently the role of these form factors is to parametrize all those effects which are not included in the theory, namely any intrinsic structure of the constituent quarks but also the quark-antiquark pair or meson cloud contributions. In order to investigate the role of the constituent quark structure, specially at high $Q^2$ \cite{wp} it would be necessary to separate the two effects. An important breakthrough in this direction is provided by a recent work \cite{sb1,bs,sb2,sbf,bfs}, where an unquenched constituent quark model for baryons has been formulated and the quark-antiquark pair contributions to the spin and the flavor asymmetry in the proton have been calculated consistently. 

The way is now open for a fully relativistic calculation of the helicity amplitudes in an unquenched CQM and a systematic study of the influence of the CQ intrinsic structure. However, the calculations presented in this paper show that, nevertheless, the CQM with a hypercentral interaction provides a reasonable basis for an overall description of the helicity amplitudes.

\section*{Appendix A. The baryon states}
\setcounter{equation}{0}

The baryon states are superpositions of $SU(6)-$configurations, which, according to Eq.
(\ref{eq:psi_tot}), can be factorized as follows:
\begin{equation}\label{eq:psi_tot1} 
\Psi_{3q}÷=÷\theta_{colour}÷÷÷ \cdot \chi_{spin}÷÷÷ \cdot \Phi_{isospin}÷÷÷
   \cdot \psi_{3q}(\vec{\rho},\vec{\lambda}).
\end{equation}

As already mentioned in the text, the various parts must be combined in order to have a
completely antisymmetric three-quark wave function. To this end it is necessary to study
the behaviour of the different factors with respect to the permutations of three objects
(that is with respect to the group $S_3$). In general, any three particle wave function
 belongs to one of the following symmetry types: antisymmetry (A), symmetry (S), mixed
symmetry with symmetric pair (MS) and mixed symmetry with antisymmetric pair (MA).

For the colour part $\theta_{colour}$ one must choose the antisymmetric colour singlet
combination.

The three-quark spin states are defined as:

\begin{equation}
\chi_{MS}~=~|((\frac{1}{2},\frac{1}{2})1,\frac{1}{2})\frac{1}{2}>,\nonumber
\end{equation}
\begin{equation}
\chi_{MA}~=~|((\frac{1}{2},\frac{1}{2})0,\frac{1}{2})\frac{1}{2}>,\nonumber
\end{equation} 
\begin{equation}
\chi_{S}~=~|((\frac{1}{2},\frac{1}{2})1,\frac{1}{2})\frac{3}{2}>,\nonumber
\end{equation} 

The antisymmetric combination is absent because there are only two states at disposal
for three particles.

Similarly one can define the isospin states $\phi_{MS}, \phi_{MA}, \phi_{S}$.

If the interaction is spin and isospin (flavour) independent, one has to introduce
products of $\chi-$ and $\phi-$ states with definite $S_3-$ symmetry. Here we give the
explicit forms only for the case that both factors have mixed symmetry, the remaining
ones being trivial:

\begin{equation}
\Omega_S~=~\frac{1}{\sqrt{2}}~[\chi_{MA} \phi_{MA} + \chi_{MS} \phi_{MS}],
\nonumber
\end{equation}
\begin{equation}
\Omega_{MS}~=~\frac{1}{\sqrt{2}}~[\chi_{MA} \phi_{MA} - \chi_{MS} \phi_{MS}],
\nonumber
\end{equation}
\begin{equation}
\Omega_{MA}~=~\frac{1}{\sqrt{2}}~[\chi_{MA} \phi_{MS} + \chi_{MS} \phi_{MA}],
\nonumber
\end{equation}
\begin{equation}
\Omega_A~=~\frac{1}{\sqrt{2}}~[\chi_{MA} \phi_{MS} - \chi_{MS} \phi_{MA}],
\nonumber
\end{equation} \label{om}

The symmetry properties of the space wave function
\begin{equation}
\psi_{3q}(\vec{\rho},\vec{\lambda})~=
\psi_{\nu\gamma}(x)~~
{Y}_{[{\gamma}]l_{\rho}l_{\lambda}}({\Omega}_{\rho},{\Omega}_{\lambda},\xi)
\label{eq:psi1}
\end{equation}
is determined by the hyperspherical part only, since the hyperradius $x$ is completely
symmetric. In Table \ref{tab:conf} we report the combinations of the hyperspherical
harmonics having definite $S_3-$symmetry.

\begin{table}[h]
\caption{Combinations $(Y_{[\gamma]l_{\rho}l_{\lambda}})_{S_3}$ of 
the hyperspherical
harmonics $Y_{[\gamma]l_{\rho}l_{\lambda}}$ that have definite $S_3-$symmetry. For
simplicity of notation, in the third column we have omitted the coupling of $l_{\rho}$
and
$l_{\lambda}$  to the total orbital ngular momentum $L$. Each combination is labelled
as $L^P_{S_3}$, specifying the total orbital ngular momentum $L$, the parity $P$ and the
symmetry type $A, M, S$.
\label{tab:conf}}
\vspace{0.4cm}
\begin{center}
\begin{tabular}{ccccccc}
\hline
& & & & & & \\
$\gamma$ & & $L^P_{S_3}$ & & $(Y_{[\gamma]l_{\rho}l_{\lambda}})_{S_3}$ & &
$S_3$ \\
& & & & & & \\
\hline
\hline
& & & & & & \\
$0$ & & $0^+_S$ & & $Y_{[0]00}$ & & $S$ \\
& & & & & & \\
$1$ & & $1^-_M$ & & $Y_{[1]10}$ & & $MA$ \\
& & & & & & \\
& & $$ & & $Y_{[1]01}$ & & $MS$ \\
& & & & & & \\
$2$ & & $2^+_S$ & & $\frac{1}{\sqrt{2}}[Y_{[2]20}+Y_{[2]02}]$
& & $S$ \\
& & & & & & \\
& & $2^+_M$ & & $Y_{[2]11}$ & & $MA$ \\
& & & & & & \\
& & & & $\frac{1}{\sqrt{2}}[Y_{[2]20}-Y_{[2]02}]$ & &
$MA$ \\
& & & & & & \\
& & $1^+_A$ & & $Y_{[2]11}$ & & $A$ \\
& & & & & & \\
& & $0^+_M$ & & $Y_{[2]11}$ & & $MA$ \\
& & & & & & \\
& & & & $Y_{[2]00}$ & & $MA$ \\
& & & & & & \\
\hline
\end{tabular}
\end{center}
\end{table}

\begin{table}[t]
\caption{Three-quark states with positive parity. The second,
third and fourth  columns show the angular momentum, parity and $S_3$-symmetry,
$L^P_{S_3}$, the spin, $S$, and isospin, $T$. States are shown in the last column  and
are written in terms of  the hyperradial wave functions, $\psi_{\nu \gamma}$,
 of the hyperspherical harmonics, $(Y_{[\gamma]})_{S_3}$ of Table 
\ref{tab:conf} and of the spin and isospin states.
\label{tab:statpos}}
\vspace{0.4cm}
\begin{center}
\begin{tabular}{ccccccc}
\hline
Resonance & & $L^P_{S_3}$ & S & T & & SU(6) configurations \\ 
\hline
$P11$  & & $0^+_S$ & $\frac{1}{2}$ & $\frac{1}{2}$ & & ${\psi}_{00}~Y_{[0]00}
~\Omega_S $
\\
& & $0^+_S$ & $\frac{1}{2}$ & $\frac{1}{2}$ & & ${\psi}_{10}~  Y_{[0]00}
~ \Omega_S$ \\
& & $0^+_S$ & $\frac{1}{2}$ & $\frac{1}{2}$ & & ${\psi}_{20}~  Y_{[0]00}
~ \Omega_S$ \\
& & $0^+_M$ & $\frac{1}{2}$ & $\frac{1}{2}$ & & ${\psi}_{22}~  \frac{1}
{\sqrt{2}}~ [ Y_{[2]00} ~\Omega_{MS} +  Y_{[2]11} ~\Omega_{MA}] $ \\
& & $2^+_M$ & $\frac{3}{2}$ & $\frac{1}{2}$ & & ${\psi}_{22}  ~\frac{1}
{\sqrt{2}}~ [\frac{1}{\sqrt{2}}~ (Y_{[2]20} - Y_{[2]02 })~\phi_{MS} +  
Y_{[2]11} ~\phi_{MA} ]~\chi_S$ \\

$P13$ & & $2^+_M$ & $\frac{1}{2}$ & $\frac{1}{2}$ & & ${\psi}_{22}
~\frac{1}{\sqrt{2}}~[\frac{1}{\sqrt{2}} ~(Y_{[2]20} - Y_{[2]02 })
~\Omega_{MS} +  Y_{[2]11} ~\Omega_{MA} ] $
\\
& & $2^+_M$ & $\frac{3}{2}$ & $\frac{1}{2}$ & & ${\psi}_{22}
~  \frac{1}{\sqrt{2}}~[\frac{1}{\sqrt{2}}~ (Y_{[2]20} - Y_{[2]02 })
~\phi_{MS} +  Y_{[2]11}~ \phi_{MA} ]~\chi_S$ \\
& & $0^+_M$ & $\frac{3}{2}$ & $\frac{1}{2}$ & & ${\psi}_{22}
~\frac{1}{\sqrt{2}}~ [Y_{[2]00} ~\phi_{MS} +  Y_{[2]11} ~\phi_{MA}]~\chi_S$\\
& & $2^+_S$ & $\frac{1}{2}$ & $\frac{1}{2}$ & & ${\psi}_{22}
~\frac{1}{\sqrt{2}}~ [ Y_{[2]20} + Y_{[2]02 }] ~\Omega_S$\\

$F15$ & & $2^+_M$ & $\frac{1}{2}$ & $\frac{1}{2}$ & & ${\psi}_{22}
~ \frac{1}{\sqrt{2}}~[\frac{1}{\sqrt{2}}~ (Y_{[2]20} - Y_{[2]02 })
~\Omega_{MS} +  Y_{[2]11} ~\Omega_{MA} ] $
\\
& & $2^+_M$ & $\frac{3}{2}$ & $\frac{1}{2}$ & & ${\psi}_{22}
~\frac{1}{\sqrt{2}}~[\frac{1}{\sqrt{2}}~ (Y_{[2]20} - Y_{[2]02 })~\phi_{MS} 
+  Y_{[2]11} ~\phi_{MA} ] ~\chi_S$\\
& & $2^+_S$ & $\frac{1}{2}$ & $\frac{1}{2}$ & & ${\psi}_{22}
~\frac{1}{\sqrt{2}}~ [Y_{[2]20} + Y_{[2]02 }]~ \Omega_S$\\

$F17$  & & $2^+_M$ & $\frac{3}{2}$ & $\frac{1}{2}$ & & ${\psi}_{22}
~\frac{1}{\sqrt{2}} ~[\frac{1}{\sqrt{2}} ~(Y_{[2]20} - Y_{[2]02 })~\phi_{MS} 
+  Y_{[2]11} ~\phi_{MA} ] ~\chi_S$\\

$P31$  & & $2^+_S$ & $\frac{3}{2}$ & $\frac{3}{2}$ & & ${\psi}_{22}
~\frac{1}{\sqrt{2}}~ [ (Y_{[2]20} + Y_{[2]02 }] ~\chi_S ~\phi_S$\\
& & $0^+_M$ & $\frac{1}{2}$ & $\frac{3}{2}$ & & ${\psi}_{22}
~\frac{1}{\sqrt{2}}~ [ Y_{[2]00} ~\chi_{MS} + Y_{[2]11} ~\chi_{MA}] ~\phi_S$
\\

$P33$  & & $0^+_S$ & $\frac{3}{2}$ & $\frac{3}{2}$ & & ${\psi}_{00}~Y_{[0]00} 
~\chi_S ~\phi_S$\\
& & $0^+_S$ & $\frac{3}{2}$ & $\frac{3}{2}$ & & ${\psi}_{10}~Y_{[0]00} ~\chi_S 
~\phi_S$\\
& & $0^+_S$ & $\frac{3}{2}$ & $\frac{3}{2}$ & & ${\psi}_{20}~Y_{[0]00} ~\chi_S~ 
\phi_S$\\
 & & $2^+_S$ & $\frac{3}{2}$ & $\frac{3}{2}$ & & ${\psi}_{22}~\frac{1}{\sqrt{2}}
~ [Y_{[2]20} + Y_{[2]02 }] ~\chi_S ~\phi_S$\\
& & $2^+_M$ & $\frac{1}{2}$ & $\frac{3}{2}$ & & ${\psi}_{22}~\frac{1}{\sqrt{2}}
 ~ [\frac{1}{\sqrt{2}} ~(Y_{[2]20} - Y_{[2]02 })~\chi_{MS} +  Y_{[2]11} 
~\chi_{MA} ]~\phi_S$\\

$F35$  & & $2^+_M$ & $\frac{1}{2}$ & $\frac{3}{2}$ & & ${\psi}_{22}
~\frac{1}{\sqrt{2}} ~ [\frac{1}{\sqrt{2}}~ (Y_{[2]20} - Y_{[2]02 })~\chi_{MS} 
+  Y_{[2]11}~\chi_{MA} ] ~\phi_S$\\
 & & $2^+_S$ & $\frac{3}{2}$ & $\frac{3}{2}$ & & ${\psi}_{22}~\frac{1}{\sqrt{2}}
~ [Y_{[2]20} + Y_{[2]02 }] ~\chi_S ~\phi_S$\\

$F37$   & & $2^+_S$ & $\frac{3}{2}$ & $\frac{3}{2}$ & & ${\psi}_{22}
~\frac{1}{\sqrt{2}}~ [ Y_{[2]20} + Y_{[2]02 }] ~\chi_S ~\phi_S$\\
\hline
\end{tabular}
\end{center}
\end{table}

\begin{table}[t]
\caption{Three quark states with negative parity. Notation as in Table \ref{tab:statpos}
\label{tab:statneg}}
\vspace{0.4cm}
\begin{center}
\begin{tabular}{ccccccc}
\hline
Resonances & & $L^P_{S_3}$ & S & T & & States \\ 
\hline
$S11$  & & $1^-_M$ & $\frac{1}{2}$ & $\frac{1}{2}$ & & ${\psi}_{11}
~\frac{1}{\sqrt{2}} ~[Y_{[1]10} ~\Omega_{MA} +  Y_{[1]01} ~\Omega_{MS}]$\\
& & $1^-_M$ & $\frac{1}{2}$ & $\frac{1}{2}$ & & ${\psi}_{21}~\frac{1}{\sqrt{2}} 
~[Y_{[1]10} ~\Omega_{MA} +  Y_{[1]01}] ~\Omega_{MS}$\\
& & $1^-_M$ & $\frac{3}{2}$ & $\frac{1}{2}$ & & ${\psi}_{11}~\frac{1}{\sqrt{2}} 
~[Y_{[1]10} ~\phi_{MA} +  Y_{[1]01} ~\phi_{MS}] ~\chi_S$\\
& & $1^-_M$ & $\frac{3}{2}$ & $\frac{1}{2}$ & & ${\psi}_{21}~\frac{1}{\sqrt{2}}
~[Y_{[1]10} ~\phi_{MA} +  Y_{[1]01} ~\phi_{MS}] ~\chi_S$\\

$D13$  & & $1^-_M$ & $\frac{1}{2}$ & $\frac{1}{2}$ & & ${\psi}_{11} 
~\frac{1}{\sqrt{2}} 
~[Y_{[1]10} ~\Omega_{MA} +  Y_{[1]01} ~\Omega_{MS}]$\\
& & $1^-_M$ & $\frac{1}{2}$ & $\frac{1}{2}$ & & ${\psi}_{21}~ \frac{1}{\sqrt{2}}
~ [Y_{[1]10}~ \Omega_{MA} +  Y_{[1]01}] ~\Omega_{MS}$\\
& & $1^-_M$ & $\frac{3}{2}$ & $\frac{1}{2}$ & & ${\psi}_{11}
~\frac{1}{\sqrt{2}}~ [Y_{[1]10} ~\phi_{MA} +  Y_{[1]01}~\phi_{MS}] ~\chi_S$\\
& & $1^-_M$ & $\frac{3}{2}$ & $\frac{1}{2}$ & & ${\psi}_{21}~\frac{1}{\sqrt{2}} 
~[Y_{[1]10} ~\phi_{MA} +  Y_{[1]01} ~\phi_{MS}] ~\chi_S$\\

$D15$  & & $1^-_M$ & $\frac{3}{2}$ & $\frac{1}{2}$ & & ${\psi}_{11}
~\frac{1}{\sqrt{2}} ~[Y_{[1]10} ~\phi_{MA} +  Y_{[1]01} ~\phi_{MS}] ~\chi_S
$\\
& & $1^-_M$ & $\frac{3}{2}$ & $\frac{1}{2}$ & & ${\psi}_{21}~\frac{1}{\sqrt{2}} 
~[Y_{[1]10} ~\phi_{MA} +  Y_{[1]01} ~\phi_{MS}] ~\chi_S$\\

$S31$  & & $1^-_M$ & $\frac{1}{2}$ & $\frac{3}{2}$ & & ${\psi}_{11}~
\frac{1}{\sqrt{2}}~[Y_{[1]10} ~\chi_{MA} +  Y_{[1]01} ~\chi_{MS}] ~\phi_S$\\
& & $1^-_M$ & $\frac{1}{2}$ & $\frac{3}{2}$ & & ${\psi}_{21}~\frac{1}{\sqrt{2}} 
~[Y_{[1]10} ~\chi_{MA} +  Y_{[1]01} ~\chi_{MS}] ~\phi_S$\\

$S33$  & & $1^-_M$ & $\frac{1}{2}$ & $\frac{3}{2}$ & & ${\psi}_{11}
~\frac{1}{\sqrt{2}} ~[Y_{[1]10} ~\chi_{MA} +  Y_{[1]01} ~\chi_{MS}] ~\phi_S
$\\
& & $1^-_M$ & $\frac{1}{2}$ & $\frac{3}{2}$ & & ${\psi}_{21}
~\frac{1}{\sqrt{2}}~ [Y_{[1]10} ~\chi_{MA} +  Y_{[1]01} ~\chi_{MS}] ~\phi_S$
\\
\hline
\end{tabular}
\end{center}
\end{table}

  In Tables \ref{tab:statpos} and
\ref{tab:statneg}, we give the explicit  form of the three-quark states with 
positive and negative parity, respectively. In these Tables the hyperradial wave
functions $\psi_{\nu\gamma}$ are solutions of the hyperradial equation Eq. \ref{eq:rad};
their form depends of course on the hypercentral potential.

\end{document}